 \documentclass[CRCHIM,Unicode,manuscript]{cedram}

\title{Accounts on Synthetic Cells}

\alttitle{Comment utiliser la classe de \protect\emph{\currentjournaltitle}\\ Un fichier \LaTeX{} exemple}

\author{\firstname{The Frontiers} \middlename{of} \lastname{Life Research Network}}
\address{Research Network of the University of Bordeaux, France} 

\author{\firstname{Sébastien} \lastname{Lecommandoux}\CDRorcid{0000-0003-0465-8603}}
\address{Univ. Bordeaux, Bordeaux INP, LCPO, UMR, CNRS 5629, Pessac, France} 
\address{Institut Universitaire de France, 75005 Paris, France}

\author{\firstname{Hamid} \lastname{Kellay}\CDRorcid{0000-0002-3438-3236}}
\address{Univ. Bordeaux CNRS LOMA UMR 5798, Talence, France} 
\addressSameAs{3}{Institut Universitaire de France, 75005 Paris, France}

\author{\firstname{C. Axel} \lastname{Innis}\CDRorcid{0000-0003-3153-9490}}
\address{Univ. Bordeaux, CNRS, INSERM, ARNA, UMR 5320, U1212, Bordeaux, France} 

\author{\firstname{Carole} \lastname{Lartigue}\CDRorcid{0000-0001-5550-7579}}
\address{Univ. Bordeaux, INRAE, UMR BFP, Villenave d'Ornon, France}

\author{\firstname{Joelle} \lastname{Amédée}\CDRorcid{0000-0002-4888-0129}}
\address{Univ. Bordeaux, INSERM, BIOTIS, UMR1026, Bordeaux, France}

\author{\firstname{Rosanna} \lastname{Le Scouarnec}\CDRorcid{0009-0009-4715-5644}}
\addressSameAs{2}{Univ. Bordeaux, Bordeaux INP, LCPO, UMR, CNRS 5629, Pessac, France}
\addressSameAs{5}{Univ. Bordeaux, CNRS, INSERM, ARNA, UMR 5320, U1212, Bordeaux, France}

\author{\firstname{Léonie} \lastname{Beaupère}\CDRorcid{0000-xxx-0000-xxxx}}
\address{Univ. Bordeaux, CNRS, CBMN, Talence, France} 
\address{Univ. Bordeaux, CNRS, IBGC, Bordeaux, France} 

\author{\firstname{Kevin} \lastname{Peyraud-Vicré}\CDRorcid{0009-0002-2367-6786}}
\address{Univ. Bordeaux, CNRS, CRPP, UMR5031, Pessac, 33600, France} 
\addressSameAs{5}{Univ. Bordeaux, CNRS, INSERM, ARNA, UMR 5320, U1212, Bordeaux, France}

\author{\firstname{Paramita} \lastname{Chaudhuri}\CDRorcid{0000-xxx-0000-xxxx}}
\address{Univ. Bordeaux, CNRS, MFP, Bordeaux, France} 
\addressSameAs{5}{Univ. Bordeaux, CNRS, INSERM, ARNA, UMR 5320, U1212, Bordeaux, France}

\author{\firstname{Seamoon} \lastname{Deb}\CDRorcid{0009-0003-4617-6378}}
\addressSameAs{8}{Univ. Bordeaux, CNRS, CBMN, Talence, France} 
\addressSameAs{9}{Univ. Bordeaux, CNRS, IBGC, Bordeaux, France}

\author{\firstname{Jitra-Marie} \lastname{Jittasevi}\CDRorcid{0000-xxx-0000-xxxx}}
\addressSameAs{6}{Univ. Bordeaux, INRAE, UMR BFP, Villenave d'Ornon, France}

\author{\firstname{Chloé} \lastname{Manseau}\CDRorcid{0009-0001-2057-5771}}
\addressSameAs{2}{Univ. Bordeaux, Bordeaux INP, LCPO, UMR, CNRS 5629, Pessac, France}
\author{\firstname{Eloise} \lastname{Equy}\CDRorcid{0009-0001-7986-7437}}
\addressSameAs{2}{Univ. Bordeaux, Bordeaux INP, LCPO, UMR, CNRS 5629, Pessac, France}
\author{\firstname{Colin} \lastname{Bonduelle}\CDRorcid{0000-0002-7213-7861}}
\addressSameAs{2}{Univ. Bordeaux, Bordeaux INP, LCPO, UMR, CNRS 5629, Pessac, France}
\author{\firstname{Emmanuel} \lastname{Ibarboure}\CDRorcid{0000-0001-8614-3851}}
\addressSameAs{2}{Univ. Bordeaux, Bordeaux INP, LCPO, UMR, CNRS 5629, Pessac, France}
\author{\firstname{Elisabeth} \lastname{Garanger}\CDRorcid{ 0000-0001-9130-8286}}
\addressSameAs{2}{Univ. Bordeaux, Bordeaux INP, LCPO, UMR, CNRS 5629, Pessac, France}

\author{\firstname{Thomas} \lastname{Salez}\CDRorcid{0000-0001-6111-8721}}
\addressSameAs{4}{Univ. Bordeaux CNRS, LOMA UMR 5798, Talence, France} 
\address{Ecole Polytechnique, Mechanics Department, Institut Polytechnique de Paris, 91128 Palaiseau, France}
\author{\firstname{Vandana} \lastname{Sharma}\CDRorcid{0000-xxx-0000-xxxx}}
\addressSameAs{4}{Univ. Bordeaux CNRS, LOMA UMR 5798, Talence, France} 
\author{\firstname{Antoine} \lastname{Aubret}\CDRorcid{ 0000-0002-4079-8104}}
\addressSameAs{4}{Univ. Bordeaux CNRS, LOMA UMR 5798, Talence, France} 
\author{\firstname{Ulysse} \lastname{Delabre}\CDRorcid{ 0000-0001-6453-3795}}
\addressSameAs{4}{Univ. Bordeaux CNRS, LOMA UMR 5798, Talence, France} 

\author{\firstname{Valérie} \lastname{Desvergnes}\CDRorcid{ 0000-0003-0290-8875}}
\address{Univ. Bordeaux, CNRS, ISM, UMR 5255, Talence, France} 
\addressSameAs{6}{Univ. Bordeaux, CNRS, INSERM, ARNA, UMR 5320, U1212, Bordeaux, France}
\author{\firstname{Jeanne} \lastname{Leblond Chain}\CDRorcid{0000-0001-8390-4811}}
\addressSameAs{6}{Univ. Bordeaux, CNRS, INSERM, ARNA, UMR 5320, U1212, Bordeaux, France}
\author{\firstname{Mikayel} \lastname{Aznauryan}\CDRorcid{0000-0002-5395-3441}}
\addressSameAs{6}{Univ. Bordeaux, CNRS, INSERM, ARNA, UMR 5320, U1212, Bordeaux, France}

\author{\firstname{Stéphane} \lastname{Arbault}\CDRorcid{0000-0002-4994-2213}}
\addressSameAs{9}{Univ. Bordeaux, CNRS, CBMN, Talence, France} 
\author{\firstname{Antoine} \lastname{Loquet}\CDRorcid{0000-0001-7176-7813}}
\addressSameAs{9}{Univ. Bordeaux, CNRS, CBMN, Talence, France} 
\author{\firstname{Yann} \lastname{Fichou}\CDRorcid{0000-0002-6520-0041}}
\addressSameAs{9}{Univ. Bordeaux, CNRS, CBMN, Talence, France} 

\author{\firstname{Anne} \lastname{Devin}\CDRorcid{0000-xxx-0000-xxxx}}
\addressSameAs{10}{Univ. Bordeaux, CNRS, IBGC, Bordeaux, France}
\author{\firstname{Patrick} \lastname{Paumard}\CDRorcid{0000-xxx-0000-xxxx}}
\addressSameAs{10}{Univ. Bordeaux, CNRS, IBGC, Bordeaux, France}
\author{\firstname{Sven J.} \lastname{Saupe}\CDRorcid{0000-0002-6236-3128}}
\addressSameAs{10}{Univ. Bordeaux, CNRS, IBGC, Bordeaux, France}
\author{\firstname{Amine} \lastname{Maarouf}\CDRorcid{0000-xxx-0000-xxxx}}
\addressSameAs{10}{Univ. Bordeaux, CNRS, IBGC, Bordeaux, France}
\author{\firstname{Damien} \lastname{Coudreuse}\CDRorcid{0000-0003-2534-1621}}
\addressSameAs{10}{Univ. Bordeaux, CNRS, IBGC, Bordeaux, France}

\author{\firstname{Vivien} \lastname{Willems}\CDRorcid{0009-0000-2634-8134}}
\addressSameAs{11}{Univ. Bordeaux, CNRS, CRPP, UMR5031, Pessac, 33600, France}
\author{\firstname{Zi} \lastname{Lin}\CDRorcid{ 0009-0002-2347-9504}}
\addressSameAs{11}{Univ. Bordeaux, CNRS, CRPP, UMR5031, Pessac, 33600, France}
\author{\firstname{Eleni} \lastname{Kopsini}\CDRorcid{0000-xxx-0000-xxxx}}
\addressSameAs{11}{Univ. Bordeaux, CNRS, CRPP, UMR5031, Pessac, 33600, France}
\author{\firstname{Esra} \lastname{Baydar}\CDRorcid{0000-xxx-0000-xxxx}}
\addressSameAs{11}{Univ. Bordeaux, CNRS, CRPP, UMR5031, Pessac, 33600, France}
\author{\firstname{Jose} \lastname{Fojo}\CDRorcid{0000-xxx-0000-xxxx}}
\addressSameAs{11}{Univ. Bordeaux, CNRS, CRPP, UMR5031, Pessac, 33600, France}
\author{\firstname{Emmanuel} \lastname{Joseph}\CDRorcid{0000-xxx-0000-xxxx}}
\addressSameAs{11}{Univ. Bordeaux, CNRS, CRPP, UMR5031, Pessac, 33600, France}
\author{\firstname{Cristian} \lastname{Villalobos-Concha}\CDRorcid{0009-0003-1212-3665}}
\addressSameAs{11}{Univ. Bordeaux, CNRS, CRPP, UMR5031, Pessac, 33600, France}
\author{\firstname{Laura} \lastname{Alvarez}\CDRorcid{0000-0001-7018-7817}}
\addressSameAs{11}{Univ. Bordeaux, CNRS, CRPP, UMR5031, Pessac, 33600, France}
\author{\firstname{Jacques} \lastname{Leng}\CDRorcid{0000-0002-7902-2442}}
\addressSameAs{11}{Univ. Bordeaux, CNRS, CRPP, UMR5031, Pessac, 33600, France}
\author{\firstname{Thomas} \lastname{Beneyton}\CDRorcid{0000-0003-2322-4886}}
\addressSameAs{11}{Univ. Bordeaux, CNRS, CRPP, UMR5031, Pessac, 33600, France}
\author{\firstname{Etienne} \lastname{Ducrot}\CDRorcid{0000-0002-1372-1144}}
\addressSameAs{11}{Univ. Bordeaux, CNRS, CRPP, UMR5031, Pessac, 33600, France}
\author{\firstname{Aurore} \lastname{Dupin}\CDRorcid{0000-0002-1372-1144}}
\addressSameAs{11}{Univ. Bordeaux, CNRS, CRPP, UMR5031, Pessac, 33600, France}
\author{\firstname{Nicolas} \lastname{Martin}\CDRorcid{0000-0003-1367-4330}}
\addressSameAs{11}{Univ. Bordeaux, CNRS, CRPP, UMR5031, Pessac, 33600, France}

\author{\firstname{Hugo} \lastname{De Oliveira}\CDRorcid{0000-xxx-0000-xxxx}}
\addressSameAs{8}{Univ. Bordeaux, INSERM, BIOTIS, UMR1026, Bordeaux, France}

\author{\firstname{Harald} \lastname{Wodrich}\CDRorcid{0000-0002-4764-1708}}
\addressSameAs{12}{Univ. Bordeaux, CNRS, MFP, Bordeaux, France} 

\author{\firstname{Bastien} \lastname{Lambert}\CDRorcid{0000-xxx-0000-xxxx}}
\addressSameAs{7}{Univ. Bordeaux, INRAE, UMR BFP, Villenave d'Ornon, France}
\addressSameAs{11}{Univ. Bordeaux, CNRS, CRPP, UMR5031, Pessac, 33600, France}
\author{\firstname{Laure} \lastname{Beven}\CDRorcid{0000-0003-4407-3255}}
\addressSameAs{7}{Univ. Bordeaux, INRAE, UMR BFP, Villenave d'Ornon, France}

\author{\firstname{Pierre} \lastname{Nassoy}\CDRorcid{0000-0003-4615-9431}}
\address{Univ. Bordeaux, CNRS, LP2N, Talence, France}

\author{\firstname{Jean-Christophe} \lastname{Baret}\CDRorcid{0000-0003-2048-8317}\IsCorresp}
\addressSameAs{11}{Univ. Bordeaux, CNRS, CRPP, UMR5031, Pessac, 33600, France}
\addressSameAs{3}{Institut Universitaire de France, 75005 Paris, France}
\email[Coordinator of Frontiers of Life]{jean-christophe.baret@u-bordeaux.fr}

\thanks{The consortium was funded by the \textit{`Réseau de Recherche Impulsion' (RRI)} \textbf{Frontiers of Life} of the University of Bordeaux (IdEx Program). It benefited from local programs of the University of Bordeaux: \textit{Programme Chaire Junior; Programme Thèses Interdisciplinaires; Programme Recherche Interdisciplinaire Exploratoire (RIE)} as well as from third-party funders (see \ref{table:2}). We acknowledge the members of \textbf{Frontiers of Life} who contributed to the networking and scientific activities which are not represented in these accounts focused on synthetic cells. We thank Marion Mathelié-Guinlet for insightful comments on the manuscript. We acknowledge D. Da Rocha for the daily management of the network activities, as well as J.-B. Verlhac, V. Jubéra and J. Cayssol of the \textit{`Département Sciences de la Matière et du Rayonnement'} at Univ. Bordeaux for their continued support. JCB acknowledges the support of V. Gabelica in the construction of the network and thanks her for useful discussions all along.}


\COI{The authors do not work for, advise, own shares in, or receive funds from any organization that could benefit from this article, and have declared no affiliations other than their research organizations.}

\begin{abstract}
Frontiers of Life is a research network of the University of Bordeaux focused on the re-engineering of living systems, starting from the construction of a synthetic cell. In this article, we present our motivation to build a cell from scratch and the elementary building blocks we worked on over the past four years. While the final objective regarding the construction of a synthetic cell remains a long-term target, the reconstruction of living cells provides both a methodology and experimental approaches to improve our understanding of living systems with potential to become the next generation of biotechnologies. 
\end{abstract}

\begin{document}
\maketitle

\twocolumngrid


\section{Building a Synthetic Cell}

Life in Nature relies on cells. They are diverse and organize at all scales, from individuals to complex tissues and multicellular organisms. At the scale of the planet, cells and organisms modify the environment, including climate and atmospheric composition as in the Great Oxidation Event or the current anthropogenic climate change~\cite{catlingHowEarthsAtmosphere2005}. From a physiological perspective, cells assemble, proliferate, differentiate and age, maintaining the large structure of the body and managing repair processes. They play central roles in disease processes, both as targets of infections and as key components of immune responses. Beyond these biological functions, the ability of cells to interact with and transform their environment is essential to natural biochemical cycles. This remarkable capacity is harnessed in engineered systems, particularly for the conversion of raw materials and for a wide range of biotechnological applications. In this biotechnology context however, living cells can not be sufficiently controlled or engineered to be adapted to the diversity of technical needs. 

Cells are now dissected and understood with a high-level of details, with critical limitations. We still fail to determine the minimal conditions – both in terms of principles and materials – for them to function. This question is intimately related to the question of the emergence of life for which an elementary definition even remains elusive~\cite{Cleland2002, bartlettDefiningLyfeUniverse2020, schrodingerWhatLife1967} if it even makes sense to search for one anyway~\cite{macheryWhyStoppedWorrying2012}. And practically, after billions of years of evolution, the genetic information encoding their functions remains both a black box and a means to oppose the modifications imposed to them in attempts to engineer them. 

From the opposition between creationists and evolutionists in the XIX\up{th} century, it is to be concluded that life forms can not spontaneously emerge from non-living matter: all living forms – as we know them – are the offspring of past living forms. The paradigm of the XIX\up{th} century summarized by Virchow that ‘\textit{only cells make cells}’ remains currently valid~\cite{schultzRudolfVirchow2008}. Even in the most advanced up-to-date demonstration of the creation of living minimal cells~\cite{gibson2010e}, the construction of a cell bearing an artificial genome requires a transplantation in a micro-compartment obtained by overwriting the genetic information of a living cell~\cite{labroussaaTransplantationGenomesRedonner2019, lartigueCreatingBacterialStrains2009, lartigue2007b}. The current biotechnology solutions all rely on the tedious engineering of existing living systems that are somehow hacked to comply with our own needs and expectations and, to date, a cell is still needed to make cells.

Leduc in his original publication `\textit{La biologie synthétique}'~\cite{clementStephaneLeducPhysique2018} in the early XX\up{th} century stated that there must be a continuum from inanimate matter to living matter in physical and chemical terms. At one point in time, there has been a transition from non-living forms to living forms, meaning that living cells have emerged or evolved from non-living matter. When placed into the realm of physical chemistry and active or out-of-equilibrium matter, fascinating research questions emerge in which labscale models will allow to test hypotheses on the emergence of life~\cite{oparinOriginLifeOrigin1965, szostakSynthesizingLife2001, buddinghArtificialCellsSynthetic2017b}. 

\begin{figure*}[tbp]
 \includegraphics[width=0.8\linewidth]{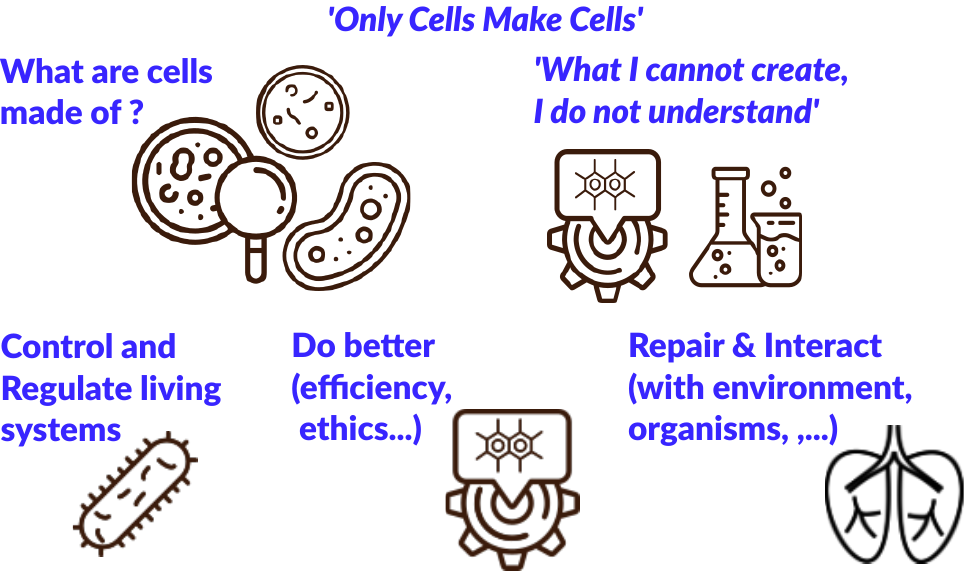}
 \caption{To challenge the Virchow principle that \textit{`only cells make cells'}, we analyse how living systems function and propose a reconstruction approach as a benchmark of our understanding of the living. In this fundamental process of understanding, we foresee that the framework generated will be useful to address problems of practical relevance to develop the next generation of biotechnologies. }
 \label{fig:1}
 \end{figure*}

A gradual implementation of such minimal systems presents a fundamental challenge with interesting engineering perspectives. A tailored assembly of new forms of living systems would be a groundbreaking step forward with massive potential for the next generation of biotechnologies. Re-engineering cells from scratch with a control of their blueprint is of interest for applications to better control and regulate living systems making use of their flaws; for applications where living cells are not efficient or not ethically acceptable; for new therapeutic solutions to tissue deficiencies. It also serves as a benchmark of our fundamental understanding of living systems, from the famous viewpoint of Richard Feynman `\textit{What I cannot construct, I do not understand}' (Figure~\ref{fig:1}).

\section{Our approach}

In Frontiers of Life, we tackle the question of the assembly of a cell from scratch, with the long-term goal to reorganize living systems from the molecular level to the level of cells, tissues and organs. As Nature does not display a cell but many forms of cells~\cite{bar-onBiomassDistributionEarth2018}, and as the essential ingredients are still not fully understood or identified to trigger life, an immense playground to construct cells from scratch is ahead of us, in which we can use the Design-Build-Test-Learn (DBTL) cycle of synthetic biology~\cite{meng2020a,kitanoSyntheticBiologyLearning2023} to construct synthetic cells. We tackle the question in an interdisciplinary approach coupling physics, chemistry, biology and engineering. 

We know what a cell is not. A cell is not simply a compartment containing a collection of molecules: it must be able to perform chemistry in an organized and sustained way, maintaining identity and carrying information over time. A cell is not a system at thermodynamic equilibrium: sustainability imposes an out-of-equilibrium condition in which energy is permanently converted as well as molecules. 

We start from the hypothesis that cellular life originated from inanimate matter through a spontaneous and gradual increase of molecular complexity~\cite{luisiEmergenceLifeChemical2006} evolving under out-of-equilibrium conditions. This complexity is based on the combination of functions that we here re-engineer. These cover: 
\begin{itemize}
 \item \textbf{Compartmentalization} with tunable bulk and membrane properties 
 \item \textbf{Energy} management and metabolism
 \item \textbf{Communication} and selective in-and-out transport
 \item \textbf{Genetic information} management
\end{itemize}

We use building blocks that we first understand and characterize and then assemble and integrate in microcompartments serving as a scaffold~\cite{schwilleMaxSynBioAvenuesCreating2018}. The building blocks can be organic or inorganic molecules and colloids, small molecules or polymers or even organelles extracted from living systems. The compartments are membrane-bound such as vesicles and polymersomes, or membrane-less, formed spontaneously by liquid-liquid phase separation~\cite{szostakSynthesizingLife2001, oparinOriginLifeOrigin1965, lin2023c, noireauxVesicleBioreactorStep2004b,palivan2024,schvartzman2024}. The assembly is based on a modular approach serving to increase complexity on demand. 

To tackle this integration, we formulate key questions related to the functions we are interested in: 
\paragraph{ \textbf{How to organize chemical systems in space and time (section~\ref{sec:3})~?}} We want to understand and mimic how the bulk phase of compartments determines its supramolecular organisation, reactivity, morphology, both under equilibrium and out-of-equilibrium conditions. Here, phase separation in liquid-liquid systems emerges as a key principle for self-organization.
\paragraph{  \textbf{How does a cell stop functioning (section~\ref{sec:4})~?} }We want to understand and mimic how the evolution of the physical chemical parameters inside a cell leads to cell death. We consider both the ageing of cellular composition revealed by its fluidity and programmable or triggered cell death mechanisms. The reconstruction approach sheds light onto the mechanism of programmable cell death at the molecular level. 
\paragraph{  \textbf{How to fuel metabolic activity using energy harvesting modules (section~\ref{sec:5})~?} } We want to mimic the processes by which cells maintain an out-of-equilibrium state by harvesting either chemical energy as in respiration or light energy as in photosynthesis. Such a module is key to sustain re-engineered biochemical processes in synthetic systems. 
\paragraph{  \textbf{How to organize membranes as responsive systems (section~\ref{sec:6})~?} } We want to design and control the membrane of the compartments to control their transport capacity. This reconstruction allows for a programmable functionalisation of the synthetic compartments, for example to chemically fuel them on demand from the outside or to activate sensing and signaling pathways.
\paragraph{  \textbf{How to program mechanical actions leading to motility (section~\ref{sec:7})~?}} We design membranes responding to external switches (temperature, electric fields) or to out-of-equilibrium forcing by the use of transducing nanomotors.
\paragraph{  \textbf{How to organize a system to boot its phenotype from its genotype (section~\ref{sec:8})~?} } We want to understand the ingredients required for a cell to boot from its genetic information, using genome transplantation as a supporting technique. Such a concept is key to further program functions in cells from genetic information. 
\paragraph{  \textbf{Our approaches make extensive use of enabling technologies (section~\ref{sec:9}).}} From a technological perspective, microfluidics is key for the manipulation, assembly and analysis of synthetic cells and for the creation and functionalization of compartments, 3d bioprinting allows for the engineering of complex environments and DNA nanotechnologies provides methods for genome transplantation and cell-free expression systems.

\section{Phase separation impacts reactivity and regulation mechanisms} ~\label{sec:3}

Phase separation is a key physicochemical process, which in cells allows for the spatial and temporal organization of chemical reactions. Emerging research suggests that the formation of biomolecular condensates, mediated by phase separation of a wide range of molecules (intrinsically disordered proteins (IDPs), nucleic acids, polyelectrolytes\dots ) is essential for regulating fundamental physiological processes up to the point where it impacts many pathologies~\cite{banani2017a,alberti2021}. Liquid-liquid phase separation is therefore a useful principle to understand~\cite{gouveia2022} and use in engineered synthetic cells~\cite{lin2023c}. 

\subsection{New chemistry assisted by liquid-liquid phase separation}

\begin{figure}[tbp]
 \includegraphics[width=\linewidth]{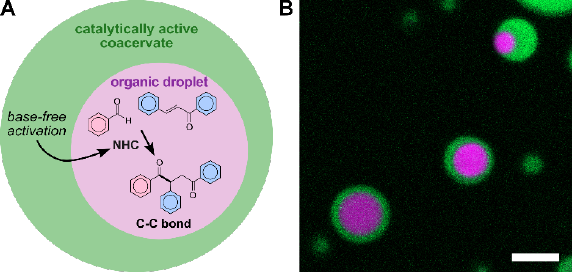}
 \caption{\textbf{New chemistry assisted by liquid-liquid phase separation}: A, In a mutliphase system, the Stetter reaction is favored under aqueous conditions, with enhanced conversion and selectivity compared with homogeneous reference conditions. B, Confocal fluorescence microscopy of the organic-in-coacervate organization. Scale bar, 10 µm. Reprinted with permission from~\cite{peyraud-vicre2025}.}
 \label{fig:2}
 \end{figure}

In cells, chemical transformations are of course under the control of thermodynamics but kinetically organized by enzymes and co-factors: catalytic activity is determined by the dynamics of their macromolecular scaffolds. The widespread occurrence of cofactors in modern metabolism allows for unfavorable reactions and it has motivated hypotheses that some cofactor-dependent chemistries may represent remnants of, or may have contributed to, early metabolic chemistry~\cite{dherbassy2024}. 

Among these cofactors, thiamine diphosphate (ThDP) is particularly intriguing because its thiazolium moiety enables C-C bond-forming chemistry through N-heterocyclic carbene (NHC) intermediates~\cite{Kern1997}. Cofactor-inspired NHC organocatalysts therefore provide a minimal chemical system to explore catalytic functions associated with the cofactors. Understanding how C-C bond formation, a key chemical reaction, emerges in minimal systems is relevant both to the construction of synthetic cells and to questions concerning the emergence of primitive metabolism.

A key question is whether a simple compartment can provide part of the physicochemical organization required for such a chemistry. Coacervates, formed by liquid-liquid phase separation of oppositely charged polymers, offer a particularly simple form of membraneless compartments. They can selectively concentrate molecular species and generate physicochemical environments that differ from the surrounding aqueous phase. This makes them relevant both as models for primitive compartmentalization and as building blocks for synthetic cells and protocellular systems~\cite{lin2023c}.

We investigated whether coacervates formed from poly(sodium acrylate) and poly(diallyldimethylammonium chloride) could organize and promote cofactor-inspired catalysis in water~\cite{peyraud-vicre2025}. We focused on the Stetter reaction between a model aldehyde and a Michael acceptor, a C-C bond-forming reaction catalysed by NHCs~\cite{debiais2020}. In the reaction conditions, the organic substrates form an immiscible organic phase that is sequestered within coacervate droplets, bringing the organic reactants into contact with the coacervate phase (Figure~\ref{fig:2}). This multiphase organization creates a reaction environment in which a newly designed amphiphilic thiazolium precatalyst tightly interacts with the polyanion-rich coacervate.

The resulting multiphase system promoted the Stetter reaction under aqueous conditions, and, importantly, without addition of an external base, with enhanced conversion and selectivity compared with homogeneous reference conditions. The coacervate was not simply acting as a compartment that sequestered the organic reactant phase. Electrostatic association between the thiazolium precursor and the polyanion favored formation of the NHC \textit{in situ}, presumably at the coacervate-organic phase interface, providing a mechanism for catalyst activation within the phase-separated system.

These results illustrate how coacervates may contribute to catalyst activation through specific molecular interactions while simultaneously organizing the reaction components within a multiphase aqueous environment~\cite{peyraud-vicrePHD}. This provides a simple example of how compartmentalization and reactivity become coupled, a principle that is central to the organization of chemistry in living cells. More broadly, cofactor-inspired catalysis in self-assembled compartments offers a conceptual link between primitive chemistry and synthetic cell engineering. Rather than viewing a compartment solely as a container into which an already functional biochemical system is introduced, this approach suggests that chemical functionality can emerge from the coupling between molecular self-assembly, compartmentalization, and reactivity. Such systems offer a route to investigate how increasingly complex chemical functions could be built from relatively simple molecular components, while providing new strategies for engineering chemically active synthetic cells.

\subsection{Phase separation with intrinsically disordered proteins}

\begin{figure}[tbp]
 \includegraphics[width=\linewidth]{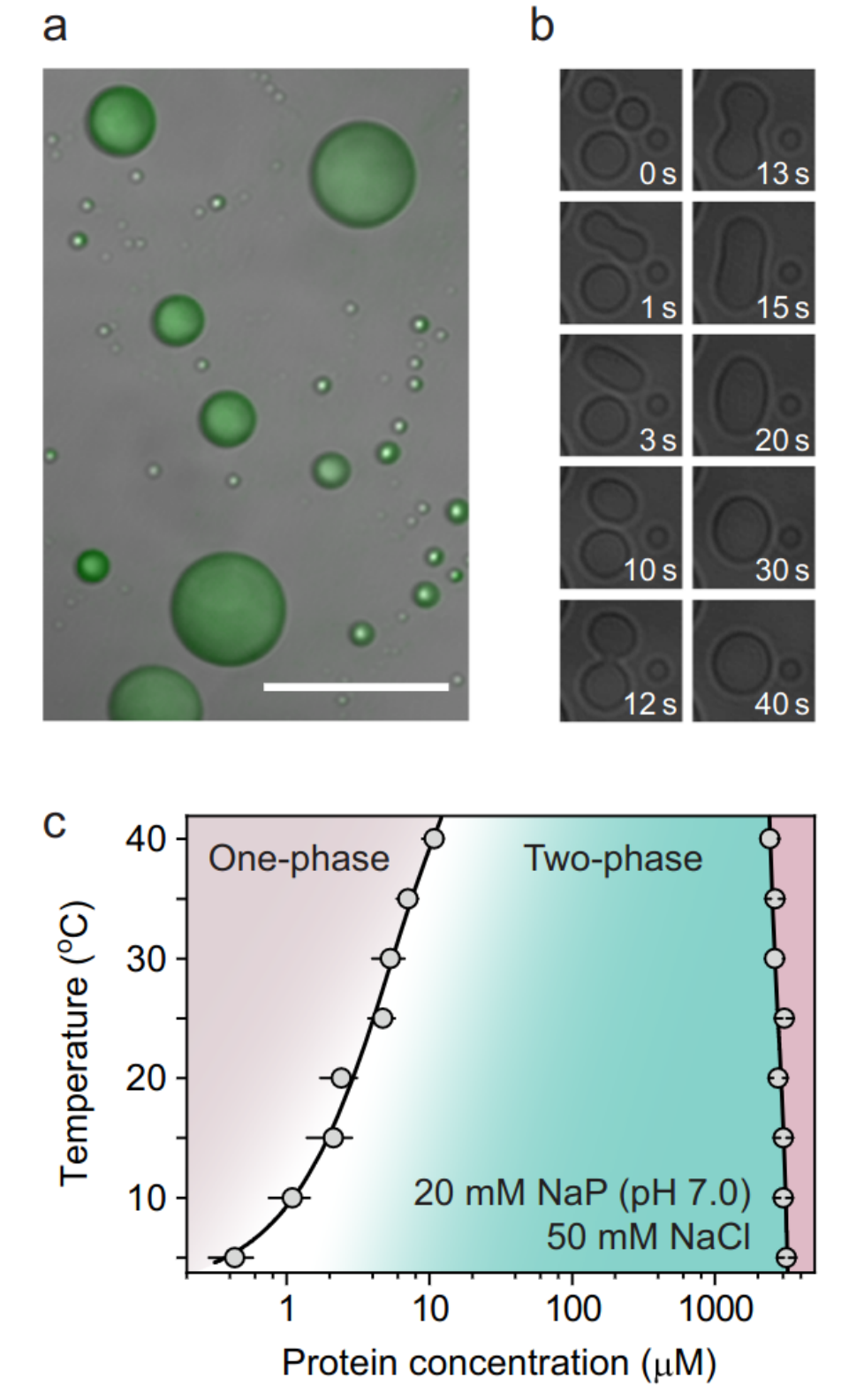}
 \caption{\textbf{Reconstruction of phase separation with intrinsically disordered proteins}. (a) Differential Interference Contrast and fluorescence images of liquid-like droplets forming upon phase separation. Scale bar: 20 microns. (b) A time course of DIC images showing rapid fusion of protein droplets. (c) Phase diagram showing the boundaries between the one-phase (purple) and two-phase (cyan) regimes. (reprinted with permission from~\cite{swain2024})}
 \label{fig:3}
 \end{figure}

Understanding the molecular mechanisms that drive condensate formation requires investigation both at the mesoscale, a regime where condensation occurs, and at the nanoscale, where the underlying molecular interactions take place. Such studies must ultimately extend from \textit{in vitro} systems to living cells.

Here we focused on \textit{in vitro} and in-cell characterization of condensation of the eukaryotic translation factor eIF4B, an intrinsically disordered translation factor implicated in regulation of protein synthesis and in the formation of stress granules-protein condensates that assemble in response to cellular stress.

Using conventional cellular imaging techniques, we confirmed that eIF4B consistently localizes within stress granules under cellular stress conditions. Moreover, eIF4B forms condensates even in cells lacking specific proteins responsible for stress granule assembly (such as G3BP1 and G3BP2). These observations suggest that eIF4B plays a significant role in stress granule biology.

Recent studies demonstrate that some proteins may self-assemble into smaller nanoscopic clusters even before the condensate formation and those nanoclusters may act as precursors for condensate formation~\cite{swain2024,kar2022}. eIF4B is one of such proteins that forms nanoclusters mediated by one of its intrinsically disordered regions heavily enriched with tyrosine residues (Figure~\ref{fig:3}). 

Through a combination of \textit{in vitro} experiments and live-cell imaging, we found that site-directed mutation of these tyrosine residues to serines abolishes eIF4B nanocluster formation and markedly reduces both \textit{in vitro} condensate formation and eIF4B recruitment to stress granules. This observation strongly points toward the key role of protein nanoclustering in promoting and regulating protein condensation \textit{in vitro} and in cells. We also explored how phosphorylation - a common chemical `switch' used by cells to control protein activity - affects condensation behavior of eIF4B. Our results show that a phosphomimetic mutant of eIF4B exhibits altered clustering and overall weaker phase separation propensity, suggesting that cells may actively use this mechanism to regulate stress granule assembly. Parallel single-molecule experiments are providing further insights into how eIF4B and its mutants interact with RNA, helping to build a comprehensive picture of the determinants governing eIF4B condensation and its localization into stress granules~\cite{chaudhuriPHD}.

\subsection{Phase separation as an out-of-equilibrium process}

\begin{figure*}[!ht]
 \includegraphics[width=\linewidth]{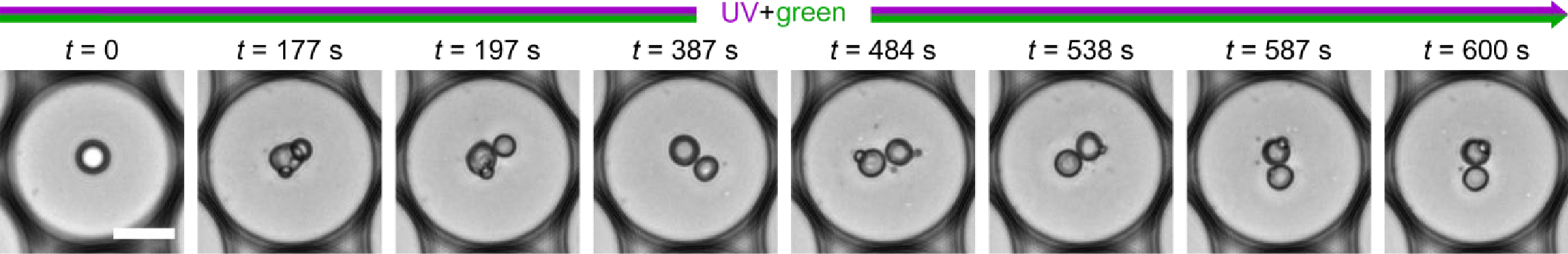}
 \caption{\textbf{Droplet division as a result of out-of-equilibrium control of liquid-liquid phase separation}. Two same-size daughter droplets persist for hundreds of seconds after division. Scale bar: 20 microns (reprinted with permission from~\cite{lin2026})}
 \label{fig:9}
 \end{figure*}

Phase separation serves as a fundamental organizing principle of soft matter, governed at equilibrium, by molecular interactions. Phase-separated droplets relax toward equilibrium by reducing their interfacial area and coarsening, ultimately yielding a single domain. When phase separation is coupled to chemical reactions, however, sustained molecular fluxes continuously compete with thermodynamic relaxation, theoretically leading to phase behaviors in ways inaccessible at equilibrium: dissipative assembly, arrested coarsening, interfacial instabilities, and division-like dynamics. We experimentally modeled a sustained molecular driving by feeding a chemically-reversible reaction with a physical energy reservoir that cannot be depleted, thus enabling the selection of non-equilibrium droplet states that can survive indefinitely using photons as an inexhaustible source of energy. 

Using a minimal light-responsive coacervate system formed by a cationic azobenzene photoswitch and DNA confined within monodisperse microfluidic droplets, we quantitatively link reaction kinetics, phase boundaries, and droplet dynamics under continuous, spatially uniform irradiation. We unravel the conditions by which tuning the relative irradiation intensities selects distinct dynamical states, from stable droplets and interfacial undulations to budding and, at higher phase stability, complete droplet division and recurrent growth-division cycles (Figure~\ref{fig:9}).

These results establish a route to select distinct non-equilibrium states in phase-separated matter, from arrested coarsening to recurrent division. They provide an experimental modeling playground to study the impact of out-of-equilibrium conditions coupled to phase separation of relevance both to understand the role of phase separation in living systems and to modulate the activity and response of synthetic cells.

\section{The limits to a sustained activity} ~\label{sec:4}

An essential step in the design and engineering of synthetic cells is to establish the physical chemistry of the intracellular space that will promote the reactions that are critical for sustaining life. A fundamental understanding and control of the mechanisms that lead to cell death is required to construct synthetic environments that have sustain activity over sufficiently long periods of time.

\subsection{Understanding the impact of physical chemistry on cellular lifespan}

We first approach this question by exploring the interplay between cellular lifespan and the biophysics of the cyto- and nucleoplasm in a model unicellular organism. In particular, parameters such as macromolecular crowding, density or diffusion rates, which we globally refer to as intracellular fluidity, are critical for governing cellular reactions both qualitatively and quantitatively. To elucidate the link between intracellular fluidity and essential cellular processes, we explore how changes in internal biophysical parameters affect cellular lifespan, using chronological aging in fission yeast cells as a model.

Chronological aging is a widespread process which is defined by the capacity of cells to re-enter the cell cycle after spending prolonged periods of time in a non-dividing, quiescent state~\cite{lopez-otin2013}. It is in fact the most common cellular state in nature and, importantly, it relies on fundamental cell functions without implicating more complex, proliferation-related processes. Previous studies using both fission and budding yeasts demonstrated that cell cycle exit and quiescence entry are accompanied by a drastic reduction in cytoplasmic fluidity and macromolecule diffusion~\cite{joyner2016,munder2016}. However, we still do not know whether these biophysical changes directly contribute to cellular lifespan, what are their quantitative limits that may contribute to the loss of cell viability, how they are modulated, and, if so, the mechanisms by which they reduce cellular lifetime.

\begin{figure*}[!ht]
 \includegraphics[width=\linewidth]{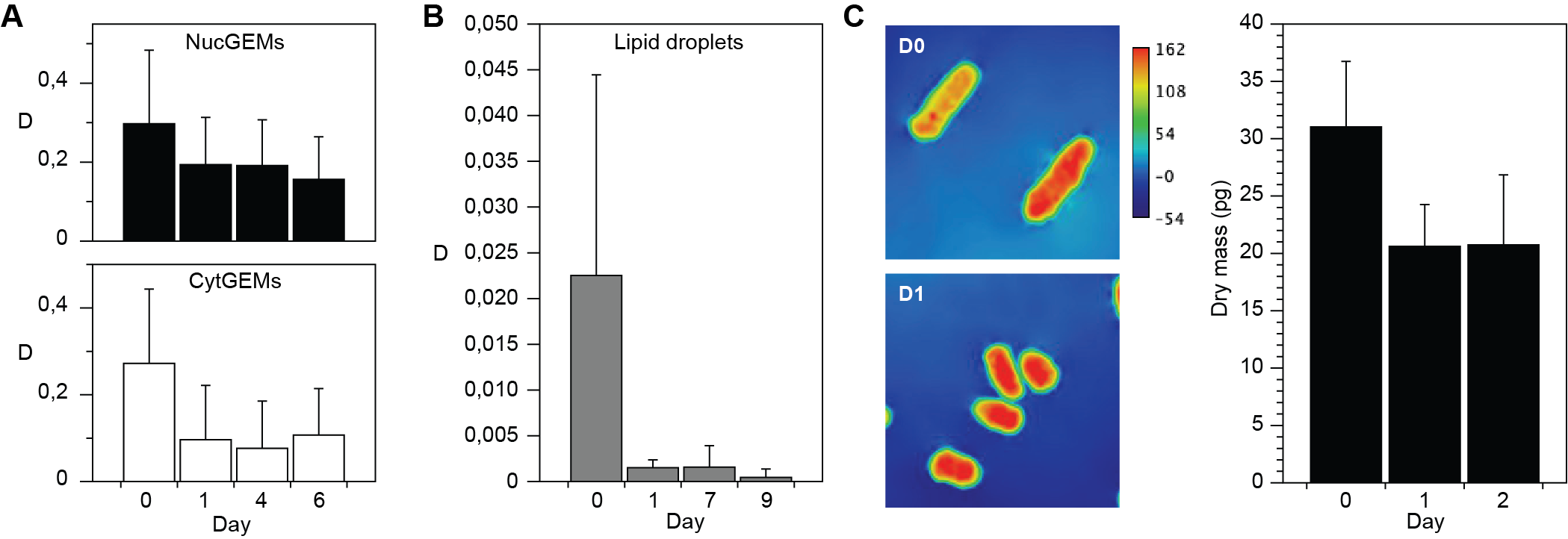}
 \caption{\textbf{Physical Chemical properties control of cell cycle}. A, B. Mean diffusion coefficients (D) of nuclear GEMs (NucGEMs, A, top panel, n>190 trajectories), cytoplasmic GEMs (CytGEMs, A, bottom panel, n>8000 trajectories) and lipid droplets (n>320 trajectories, B) in fission yeast cells at different days after cell cycle exit (Day 0: exponential growth). The diffusion coefficients were computed for each particle trajectory by performing a linear fit on the first three points of the mean squared displacement (MSD) curve using the relation MSD=4Dt. C. Determination of single cell dry mass by quantitative phase imaging (QPI). Left panels: representative QPI images in proliferation (D0) and after 1 day in quiescence (D1). Color scale: the intensity of each pixel is proportional to the dry mass density. Right panel: quantification of single cell dry mass at different times after cell cycle exit (Day 0: exponential growth). n=40 for each time point. Error bars are standard deviations.}
 \label{fig:13}
 \end{figure*}

The first step in our approach is to determine the set of parameters to use in order to best quantify cellular fluidity, as a single method cannot fully account for the complexity of the internal biophysical properties of a cell. We therefore implemented complementary technical and analytical pipelines to acquire and analyze live-cell imaging data for assessing fluidity at different scales. For this, we combine FRAP (Fluorescence Recovery After Photobleaching – single protein diffusion), GEM tracking~\cite{delarue2018} (Genetically Encoded Multimeric nanoparticles of 40 nm – both cytoplasmic and nuclear), lipid droplet tracking (large particles of various sizes), and Quantitative Phase Imaging (determination of dry mass in living cells) coupled to cell volume measurements to derive single-cell density. Using these tools, we establish single-cell maps of intracellular crowding and fluidity over time in non-dividing cells as they age (Figure~\ref{fig:13}). Importantly, we take advantage of both wild-type cells as well as aging mutants showing distinct aging patterns and overall lifespan~\cite{romila2021}. Correlating our measurements with the timing and capacity of the cells to resume growth upon exposure to proliferative cues establishes potential biophysical thresholds associated with irreversible quiescence and subsequent cell death. We show that the changes in intracellular fluidity that occur as cells exit the division cycle and age are complex rather than reflecting a simple and discrete biophysical transition (unpublished results). Furthermore, our data also suggest that quiescence entry and growth resumption are associated with specific quantitative fluidity targets. 

Next, we modulate the intracellular fluidity and evaluate its impact on the lifespan of quiescent, non-dividing cells. First, we identified a set of fluidity mutants, which show distinct levels of crowding in quiescence\footnote{collaboration with Dr. Karsten Weis, ETH Zurich, Switzerland}. Second, we artificially modulated fluidity through induced alteration of membrane permeability (water transport), and exposure of cells to higher osmotic pressure, leading to volume loss and cells with a long-term, sustained decrease in fluidity. Finally, by implementing an orthogonal system to trigger the formation of gas vesicles within cells ~\cite{sawyer2021}, we forced the local increases in macromolecular crowding that are directly related to the sizes of the vesicles\footnote{collaboration with Dr. Pascal Hersen, Curie Institute, France}. We are now taking advantage of these different strategies to assess the regulation of lifespan by intracellular biophysical properties, as well as the biological processes involved in a reduction vs. extension of viability and our preliminary data point toward the possibility that intracellular fluidity is a rheostat for cellular lifespan. 

Finally, the formation of irreversible aggregates is an emerging feature of aging and the accompanying decrease in viability ~\cite{hondele2020, alberti2019, joyner2016, lopez-otin2013}. Notably, this phenomenon is influenced by the concentration of macromolecules in the cyto- and nucleoplasm. We are therefore coupling the models and approaches described above with aggregation markers to determine whether intracellular fluidity also contributes to shape lifespan through governing the timing of appearance of these aggregates during the aging process as well as their properties when cells lose viability (e.g. reversibility, size). Complementary to these live-cell imaging methods, we use a mass spectrometry-based technique to characterize these aggregates~\cite{schopper2017} (e.g. properties of the aggregated proteins, charges, fraction of proteins showing intrinsically disordered regions). 

Beyond understanding the biology of aging, we provide new insight into the biochemical and physical limits of the intracellular space that need to be considered in the design of synthetic cells.

\subsection{Reconstitution of programmable cell death mechanisms}

Programmed cell death is identified since the past decades as a near-universal feature of living organisms throughout the Tree of Life, including unicellular life forms~\cite{hofmann2020, lopatina2020, maekawa2023}. These cell death processes refer to the controlled elimination of a cell through molecular mechanisms encoded by the organism itself, rather than as a passive consequence of accidental damage. Programmed and regulated cell death mechanisms play critical roles in organismal homeostasis, development and defense, including antiphage defense in bacteria. 

In a paradox that is only apparent, one is left with the realization that cell death is essential for cellular life. It can therefore be argued that both synthetic cells and living cells engineered to retain only minimal biological functions should incorporate mechanisms enabling their controlled elimination. More broadly, efforts aimed at reconstituting life-like processes could benefit from including a mechanism for the controlled elimination of a given cell population, either to mimic developmental or differentiation processes or to reproduce evolutionary bottlenecks. Such cell elimination processes could also represent an important safeguard for the potential deployment of synthetic cells in applied science. In the long run, biotechnological deployment of synthetic cells might ultimately result in the emergence of virus-like entities targeting artificial cells, making regulated-cell-death-based defense systems. 

Our long-term goal here is to incorporate a regulated cell death machinery into a protocell environment. As a biological model, we turned to protein-based regulated cell death modules found in microbes, particularly fungi and bacteria~\cite{saupe2020}. These regulated cell death modules are activated through a process termed amyloid signaling, whereby the controlled aggregation of an amyloid-forming receptor module triggers, through a process known as cross-seeding, the amyloid templating of a downstream cell death execution module that ultimately functions as a pore-forming toxin. Interestingly, these cell death execution modules can be transferred between biological systems and remain capable of inducing regulated cell death across a wide range of cell types and organisms, including unicellular and multicellular eukaryotes as well as bacteria ~\cite{ibarlosa2025}. Achieving incorporation of this machinery into protocells requires the development of a robust workflow for protein encapsulation and the design of fluorescently labeled amyloid proteins that retain their ability to self-assemble within confined protocell environments. 

\begin{figure}[tbp]
 \includegraphics[width=0.9\linewidth]{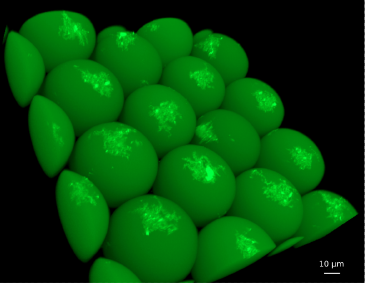}
 \caption{\textbf{Reconstitution of Programmable Cell Death Mechanisms}: amyloid proteins encapsulated in a soluble form are assembled \textit{in situ} in synthetic compartments - here droplets - leading to the formation of amyloid fibrils inside the protocells, as observed in confocal fluorescence (unpublished result).}
 \label{fig:4}
 \end{figure}

As a proof of concept, we focus on the encapsulation of amyloid proteins within synthetic protocells and their controlled molecular assembly \textit{in situ}. We first prepared several protein constructs associated with different fluorescent tags to enable their observation by confocal microscopy and developed a workflow for encapsulating amyloid proteins within protocells, taking advantage of the rapid generation of microdroplets using a microfluidics-based workflow. We demonstrated that amyloid proteins could be encapsulated in a soluble form and that their molecular assembly could subsequently be triggered \textit{in situ}, leading to the formation of amyloid fibrils inside the protocells (Figure~\ref{fig:4}).

Structural biology techniques, including solid-state NMR, confirmed that the encapsulated amyloid fibrils adopt a three-dimensional structure similar to that observed under \textit{in vitro} conditions. Finally, we demonstrated that the amyloid formed in the encapsulated environment remains bioactive by inducing the native regulated cell death cascade when reintroduced into fungal cells. Future efforts will focus on the simultaneous incorporation of the amyloid trigger and the downstream amyloid-controlled cell death execution module, to ultimately reconstitute a complete regulated cell death pathway within a protocell~\cite{debPHD}. 

Beyond the strict cell-death related aspects, our approach offers an opportunity to study and possibly fine tune amyloid-formation processes (and by extension other phase separation processes) in a confined protocell environment. From a methodological perspective, the study of amyloid assembly in microdroplets, biomimetic membranes, and microfluidic channels involving flows~\cite{audeoud2026} open vast possibilities to quantitatively address key questions in the field, including the kinetics of amyloid nucleation or homologous or heterologous amyloid seeding mechanisms.

\section{Energizing synthetic cells} ~\label{sec:5}

How can we provide the energy required to maintain a cellular system running? We need to define minimal metabolic networks able to convert energy and feed it into the system to fuel the cellular functions. One of the essential proxy to transfer energy is the conversion of cellular energy into ATP and cofactors of the metabolism~\cite{walsh2018}. The \textit{in situ} synthesis of this energy sustains biological activities (transport, synthesis, conversion, etc.) that are durable and required over the long term for cell survival and reproduction~\cite{hwang2024}. 

The engineering from scratch of an energy harvesting module is an outstanding task, but an example of the potential to reuse existing machineries has been shown: bacterial cells encapsulated in polyelectrolyte coacervate have been shown to power the cellular machinery of lysed bacteria inducing ATP drive morphological restructuring~\cite{xu2022a, xu2025}. The integration of subcellular systems have also been proposed: the use of inverted membrane vesicles allows for the regeneration of cofactor ATP with the conversion of NADH into NAD$^+$~\cite{beneyton2018c}. In this case, an out-of-equilibrium state was shown to be sustained between two limiting states, the active state of the compartment and the sleeping state~\cite{beneyton2018c} by coupling opposing reactions. Inspired by the diversity of natural cells that are either autotrophes or heterotrophes, we now provide routes to energize microcompartments, based on the integration of subcellular machineries such as mitochondria and thylakoids, for cofactor regeneration. 

\subsection{Chemical energy}

\begin{figure*}[tbp]
 \includegraphics[width=\linewidth]{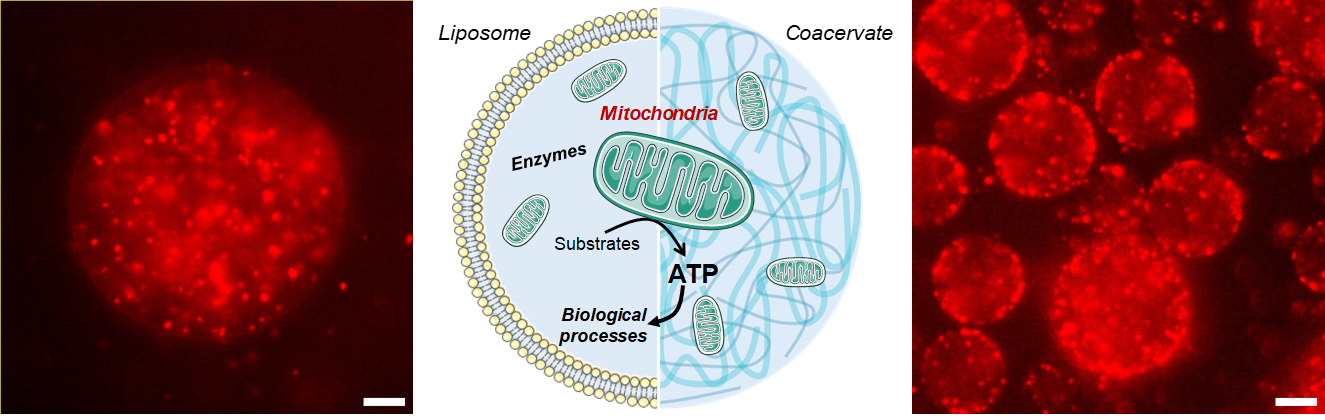}
 \caption{\textbf{Integration of mitochondria in synthetic compartments}: Schematic of a bioenergetic artificial cell incorporating mitochondria; \textit{in situ} observation of their activity (inner membrane potential) by epifluorescence microscopy when encapsulated in a giant liposome (left) or in coacervates (right). scale bar: 10 microns (unpublished results)}
 \label{fig:10}
 \end{figure*}

Among the sources of cellular bioenergy, mitochondria are the main ATP-producing organelles in the aerobic processes of eukaryotes. Here, we encapsulate mitochondria isolated from cells (\textit{Saccharomyces cerevisiae} yeast) within synthetic microcompartments to form bioenergetic artificial cells (Figure~\ref{fig:10}). The use of a specialized organelle, the mitochondrion, allows ATP to be produced in large quantities and at controlled fluxes~\cite{Bioenergetics22014, li2022}.

Two types of synthetic microcompartments were used: giant liposomes (GUVs, giant unilamellar vesicles), made of phospholipid membranes mimicking natural cell membranes, and coacervates, droplets produced by liquid–liquid phase separation of biocompatible polymers, forming membrane-free compartments with a dense, hydrated core similar to the cell cytosol. Both types of compartments are biomimetic, as they provide controlled structures and environments favorable to biochemical and biological reactions~\cite{lin2023c,nair2023}. We characterized the oxidative phosphorylation activity of the encapsulated mitochondria, at the single-organelle level and in presence or absence of substrates and inhibitors. We used epifluorescence microscopy to assess the inner membrane potential, the mitochondrial NADH content, and the production of reactive oxygen species~\cite{weissig2021,lefrancois2021}. ATP synthesis was measured by bioluminescence and electrochemical sensors were used to monitor oxygen consumption and the rate of reduction of exogenous quinones by the encapsulated mitochondria. The complexity was then increased by using substrate-regenerating enzymes to enhance the level of autonomy of the biomimetic compartment.

In giant liposomes made of phospholipids and fatty acid mixtures, we observed that encapsulating mitochondria at high concentration preserves an intact population capable of establishing a membrane potential. Energy production by encapsulated mitochondria was observed under addition of ADP, through the ATP synthesis and complete consumption of ADP monitored over time. These results provide a proof of principle that mitochondria encapsulated in liposomes can behave as a `bioenergetic power plant'. Finally, preliminary experiments were carried out to integrate peptide-based pores into the liposome membranes. This could allow passive transport of ADP across the membrane and thus enable continuous ATP synthesis.

In coacervates, the encapsulated mitochondria show variations in membrane potential and NADH levels that can be observed directly by microscopy. The phosphorylating activity was assessed by measuring oxygen consumption and ATP synthesis, with results equivalent to those obtained for free mitochondria in solution. The use of exogenous redox mediators, namely water-soluble quinones, made it possible to characterize electrochemically, remotely and with an electrode placed in the medium, the rate at which quinones are reduced by the respiratory chain of the encapsulated mitochondria. Finally, the ATP produced by oxidative phosphorylation was used to power other systems, such as a bi-enzymatic cycle (hexokinase and G6PDH) that regenerates NADH. This NADH is then used by the mitochondria as a reduced substrate, closing a loop of biochemical activities centered on ATP. Coacervates thus constitute compartments well suited to encapsulate mitochondria and carry out biochemical and bioenergetic reactions in a dense environment that mimics the intracellular medium, paving the way toward an autonomous bioenergetic artificial cell~\cite{beauperePHD}.

 \subsection{Harvesting light energy to power metabolism}
 
\begin{figure}[tbp]
 \includegraphics[width=0.9\linewidth]{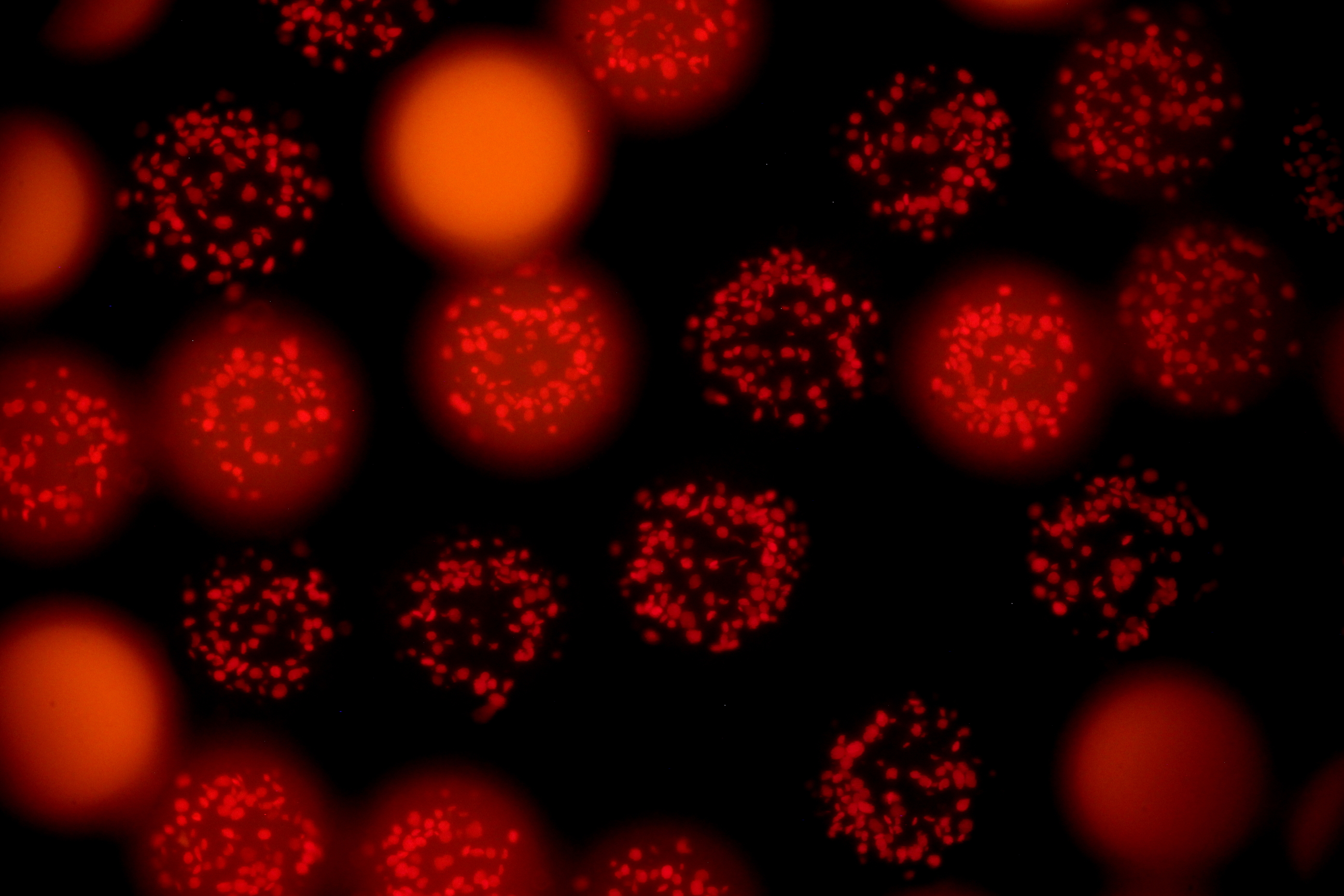}
 \caption{\textbf{Harvesting Light Energy using thylakoids in synthetic compartments}. The encapsulation of thylakoids in droplets (here in red fluorescence allows for the regeneration of ATP and NADPH is monodisperse droplets upon light illumination~\cite{miller2020a}}
 \label{fig:55}
 \end{figure}

Among the energy harvesting mechanisms of living systems, photosynthesis is the most important process to provide energy to realize thermodynamically unfavorable reactions. Miller \textit{et al.} (2020) developed a chloroplast mimic that combines the natural photosynthetic membranes of thylakoids with a synthetic CO$_2$-fixation pathway~\cite{miller2020a}, the synthetic CETCH cycle, an engineered pathway that converts CO$_2$ into organic molecules~\cite{schwander2016c} (Figure~\ref{fig:55}). Upon illumination, the photosynthetic membranes generate the energy-rich compounds ATP and NADPH required to drive the enzymatic reactions, thereby establishing a functional connection between light energy and carbon assimilation. By reconstituting these components within cell-sized compartments, a mimic of the naturally occurring energy-conversion systems is reconstructed to power synthetic metabolic networks in a controlled, cell-like environment. Using microfluidics to generate and control cell-sized droplets, millions of monodisperese, independent synthetic compartments can be produced to generate large scale populations. The composition of these compartments can be adjusted to optimize reaction conditions, while their activity can be monitored in real time through NADPH fluorescence. Furthermore, the use of light as an external stimulus offers a means of controlling metabolic activity without directly modifying the biochemical components of the system. 

This work illustrates a promising strategy for energizing synthetic cells by exploiting the natural photosynthetic machinery rather than reconstructing the entire photosynthetic apparatus from scratch. Although the chloroplast mimic does not yet constitute a fully autonomous synthetic cell, it establishes an important framework for integrating energy conversion, carbon fixation, and compartmentalization within a single artificial system. Future developments will extend these approaches to power other ATP dependent processes, such as protein synthesis or mechanical actions. 

Ultimately, combining natural energy-conversion modules with synthetic metabolic networks could help to overcome one of the major limitations of synthetic biology: the establishment of a sustained and externally-controllable energy supply for increasingly complex, life-like systems.

\section{Responsive membrane-bound compartments} ~\label{sec:6}

The membrane of a cell plays a central role in defining the identity of the cell, acting as a physical boundary and as a responsive interface with a capability to selectively control which molecules are transported through, as well as which ones are retained and when. Conceptually, in soft, crowded, charged, and fluctuating liquid-like environments, the transport dynamics of molecules, proteins and nanoparticles and their first-encounter properties to pores might differ from the ones in diluted bulk samples, with potential implications for the clock of the life machinery. As an example, using model experimental physical systems, we recently demonstrated the emergence of boundary effects affecting the probability of displacements of particles, through non-Gaussianities \cite{alexandre2023}, enhanced rare events and enhanced target finding dynamics \cite{detournemire2026}. We also revealed the emergence of intriguing novel forces and drifts associated with soft Brownian couplings to interfaces~\cite{fares2024} with ongoing efforts to study biomimetic lipidic membranes decorated with hyaluronic acid brushes~\footnote{Collaboration Aurelia Honerkamp-Smith, Lehigh Univ., USA}. Controling transport accros membranes therefore represents both a remaining fundamental question and a challenge of practical relevance in the desing of synthetic cells. 

\subsection{Switchable transport}

\begin{figure*}[tbp]
 \includegraphics[width=0.8\linewidth]{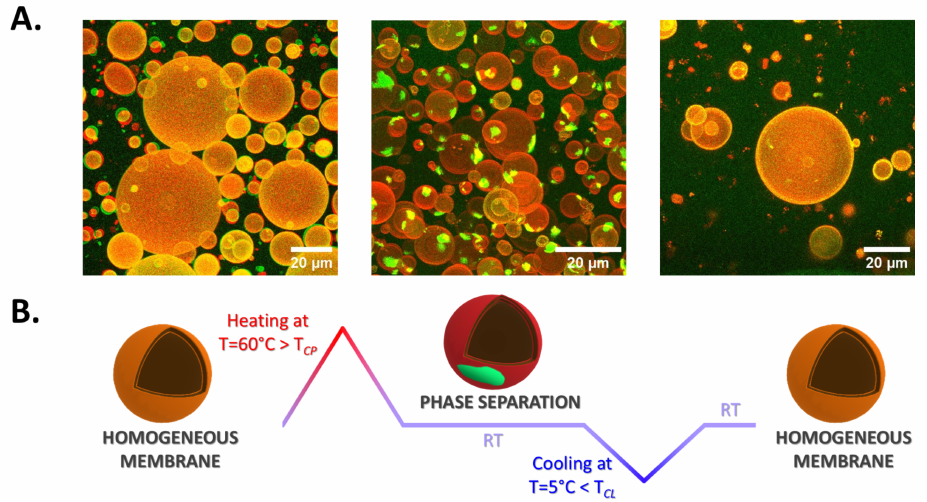}
 \caption{\textbf{Transport across membranes controlled by interfacial phase transition}: Transport across membranes is controlled by the interfacial phase transition of polyproline. Confocal stacks of vesicles (A) and the corresponding scheme (B) were performed after the electroformation of GUVs at room temperature (left) and after heating (middle). The process is reversible when the system is cooled down (right). Scale bar: 20 microns.The anchoring of polyproline chains to the membrane leads to reversible phase separation, with clusters forming at high temperature and dispersing upon cooling. The microdomains enhanced membrane permeability, enabling size-controlled release of encapsulated molecules, including synthetic polymers and enzymes. (reprinted with permission from ~\cite{lescouarnec2025})}
 \label{fig:5}
 \end{figure*}

Cells control tightly the transport of molecules across their membranes, and similarly controlled import and export is a crucial function to develop and characterize for synthetic cells. For membrane bound systems (vesicles or polymersomes), the transport processes must discriminate the key molecules that maintain the identity of the cell and encode the functional molecules from the nutrients and waste products~\cite{lu2022a}. Transmembrane transport of solutes is a critical feature to build synthetic cells and to develop therapeutic nanocarriers for nanomedicine~\cite{Zhao2022}. In Nature, the diffusion across the cellular membrane is either facilitated by proteins or by unique lipid dynamics~\cite{shin2022}.

In order to achieve improved control over permeability in cell-like compartments, synthetic systems, such as those based on polypeptides, are particularly promising~\cite{marguet2013d}. These polymers possess properties similar to proteins, including their capacity to self-assemble via well-defined secondary structures, as well as stimuli-responsive properties, biocompatibility, and biodegradability~\cite{Huang2013}. Anchoring these polymers in a cell membrane can be facilitated by lipid functionalization~\cite{izgu2016} and stimuli-responsive lipids~\cite{viricel2015a} provide strategy to destabilize membranes. A major challenge in the design of synthetic membrane-active systems is to combine structural stability with the dynamic responsiveness and reversibility of biological membranes. In this context, polyproline is an attractive synthetic polypeptide owing to its unique conformational properties and distinctive physicochemical behavior~\cite{scouarnec2026}. It exhibits an unusual thermoresponsive behavior characterized by a lower critical solution temperature (LCST) transition with pronounced hysteresis~\cite{badreldin2022} (Figure~\ref{fig:5}). Building on these properties, we developed lipid–polyproline conjugates to transfer the intrinsic thermoresponsiveness of polyproline to lipid membranes~\cite{lescouarnec2025}. By anchoring polyproline chains within the lipid bilayer, we induced temperature-dependent membrane reorganization through the reversible formation of polymer-rich microdomains~\cite{lescouarnec2025}. This process was associated with controlled changes in membrane permeability, enabling the transport and release of hydrophilic molecules ranging from small solutes to macromolecules, including enzymes~\cite{lescouarnec2025}. 

These lipid–polyproline assemblies thus provide a simplified biomimetic model sharing certain functional features with pore-forming membrane proteins, such as hemolysins, particularly their ability to modulate membrane permeability. The approach was further extended to liposomes of different sizes, highlighting its potential for both artificial cell models and nanocarrier applications~\cite{lescouarnec2026,lescouarnecPHD}.

\subsection{Cargo release through membrane phase separation and remodeling}

\begin{figure*}[tbp]
 \includegraphics[width=\linewidth]{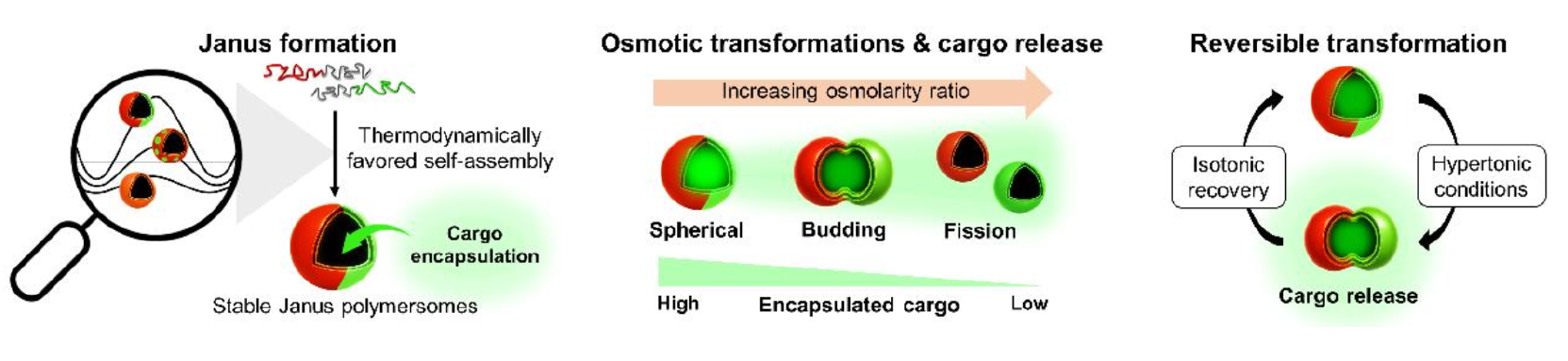}
 \caption{\textbf{Design for cargo release}: A thermodynamic framework establishes the conditions for the formation of kinetically stable Janus polymersomes using a solvent-assisted strategy. Osmotic stress subsequently drives their membrane remodeling from reversible budding to controlled fission. These dynamic morphological transformations are accompanied by progressive and tunable release of encapsulated cargo, establishing Janus polymersomes as dynamic compartments with tunable morphology and cargo release.}
 \label{fig:14}
 \end{figure*}

The ability of biological membranes to dynamically reorganize their structure and morphology is essential for many cellular functions, including compartmentalization, membrane trafficking, and cell division. In this context, polymersomes assembled from amphiphilic block copolymers offer a versatile platform for engineering membranes with tunable physicochemical and mechanical properties. In particular, Janus polymersomes, characterized by the lateral phase separation of two chemically distinct copolymers within a single membrane, provide a unique opportunity to investigate the interplay between membrane composition, morphological transformations, and controlled molecular release. We first demonstrated that membrane phase separation could be rationally controlled by tuning the chemical nature and relative incompatibility of the constituent block copolymers~\cite{equy2025}. Using a thermodynamic framework based on Flory-Huggins theory, we established a relationship between the Gibbs free energy of mixing and the resulting membrane morphology, allowing the formation of homogeneous, patchy, or Janus polymersomes to be predicted and experimentally controlled. This approach enabled the preparation of well-defined Janus giant unilamellar vesicles (JGUVs) with high yields, while revealing that excessive copolymer incompatibility could promote membrane fission and prevent the stabilization of Janus architectures. Building on these findings, we further demonstrated that the formation and stability of Janus polymersomes could be extended to a much broader range of copolymer incompatibilities using a solvent-assisted double-emulsion strategy~\cite{manseau2026}. We demonstrated that osmotic stress can trigger controlled membrane deformation, budding, and ultimately fission, with the possibility of reversing budding under appropriate conditions (Figure~\ref{fig:14}). By adjusting the osmotic conditions, the extent and dynamics of these morphological transformations can be modulated. This membrane remodeling can be exploited to achieve staged and tunable release of encapsulated cargo, as demonstrated using fluorescein as a model molecule. These results establish a direct functional connection between membrane organization, externally induced morphological transformations, and the regulation of molecular release.

\subsection{Mechanically responsive membranes}

Beyond equilibrium behaviour, living cells are intrinsically non-equilibrium systems, continuously consuming energy to maintain their organization. The dynamics of their membranes is essential in biological processes such as communication, division and motility~\cite{willems2026}.

\begin{figure}[tbp]
 \includegraphics[width=\linewidth]{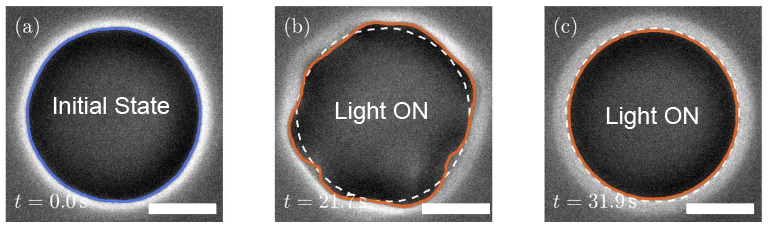}
 \caption{\textbf{Thermally induced deformation of a DOPC giant unilamellar vesicle (GUV)}: (a) Initial spherical configuration before illumination (b) Large-amplitude membrane deformation following light activation. (c) Recovery of a nearly spherical shape under continued illumination. Solid lines indicate the vesicle contours, while the dashed line in indicates the reference spherical shape. Scale bar 10 microns (unpublished data).}
 \label{fig:8}
 \end{figure}

We explore a mechanism of non-equilibrium membrane dynamics by investigating how rapid thermal perturbations can induce large-amplitude deformations and transient shape fluctuations. Temperature offers a versatile route to control membrane dynamics, as it simultaneously influences several of their physical properties. Pioneering studies in the early 1990s demonstrated that temperature-induced changes in the membrane area-to-volume ratio and bilayer asymmetry can drive vesicle shape transformations~\cite{kas1991}. Nevertheless, in this studies temperature is changed slowly compared with membrane relaxation. Rapid temperature changes influences membrane fluidity, thermal area expansion, bending rigidity, and tension. It can also drive the membrane away from mechanical equilibrium, generating transient excess area and shape instabilities as the membrane and enclosed fluid relax toward a new equilibrium configuration. To investigate these effects, we exploit a recently-developed microscope-integrated optothermal platform that uses incoherent wide-field illumination and a thin absorbing metal film to generate rapid, localized, and controllable temperature variations without the need to incorporate photosensitive molecules into the lipid bilayer~\cite{villalobos-concha2026}. We prepare DOPC vesicles via hydration, which are suspended in a sucrose-glucose solution so they sediment close to the bottom surface. The sample is illuminated under the blue light of the inverted miroscope, which leads to an increase of 20$^o$C in less than 10ms. Upon this fast temperature shock, the membrane fluctuations are strongly amplified, accompanied by a rapid increase in the projected area, which reaches a maximum within the first seconds of heating. Despite continued illumination, the initial deformation is followed by pronounced contraction of the projected contour. The large fluctuations are progressively supressed, and the vesicle approaches a nearly circular profile, with a projected area smaller than its initual value. After switching off the light, the vesicle partially recovers its initial morpholohgy and fluctuation amplitude (Figure~\ref{fig:8}\footnote{Collaboration with Dr. Andela Saric, ISTA Vienna, Austria}). 

These results provide insight into the response of lipid membranes to rapid thermal perturbations and establish temperature shocks as a route to control morphological dynamics in minimal synthetic cell-like systems.

\subsection{Membrane deformation through collective excitation of nanomotors}

Another route to produce membrane deformation is by the coupling of the compartment bulk activity: living cells composed of soft membranes are actuated by the collective action of energy-transducing nanomotors. The organized activity of these internal motors governs motility and sensing mechanisms at both the individual and collective scales. Understanding and mimicking the effects of active forces within a soft membrane remains a significant challenge~\cite{needleman2017}. For instance, a few studies have recently leveraged self-propelled microparticles to generate internal stresses, ranging from bacteria to micrometric colloids~\cite{takatori2020,vutukuri2020a}. As a next step, we aim to introduce elementary programmability into the dynamics of soft compartments driven by synthetic building blocks inside the core of the compartment. 

We investigate the coupling between the mechanics and motility of a membrane containing internalized nano-transducers. We use polymersomes as model giant unilamellar vesicles and encapsulate gold nanoparticles (AuNPs) that behave like optical nanotransducers. Under laser illumination, in addition to optical forces, AuNPs absorb light and convert it into localized heat, driving the system far from equilibrium. This induces a strong optomechanical response, revealing a coupling between both the internal and membrane dynamics. The polymersomes are formed using poly(butadiene)-b-poly(ethylene oxide) (PBut2.5-bPEO1.3) copolymers to create a soft membrane, following a double emulsion method~\cite{hayward2006,peyret2018}. AuNPs are encapsulated at an early stage of the emulsion. Excitation is performed using a 532 nm gaussian laser beam close to the plasmon resonance of AuNPs (size 100 and 300 nm). We observe the polymersomes and track their dynamics using phase-contrast microscopy, taking advantage of the slight refractive index difference between the internal (sucrose, 300 mOsm) and external (glucose, 305 mOsm) buffers of the polymersomes.

\begin{figure}[tbp]
 \includegraphics[width=\linewidth]{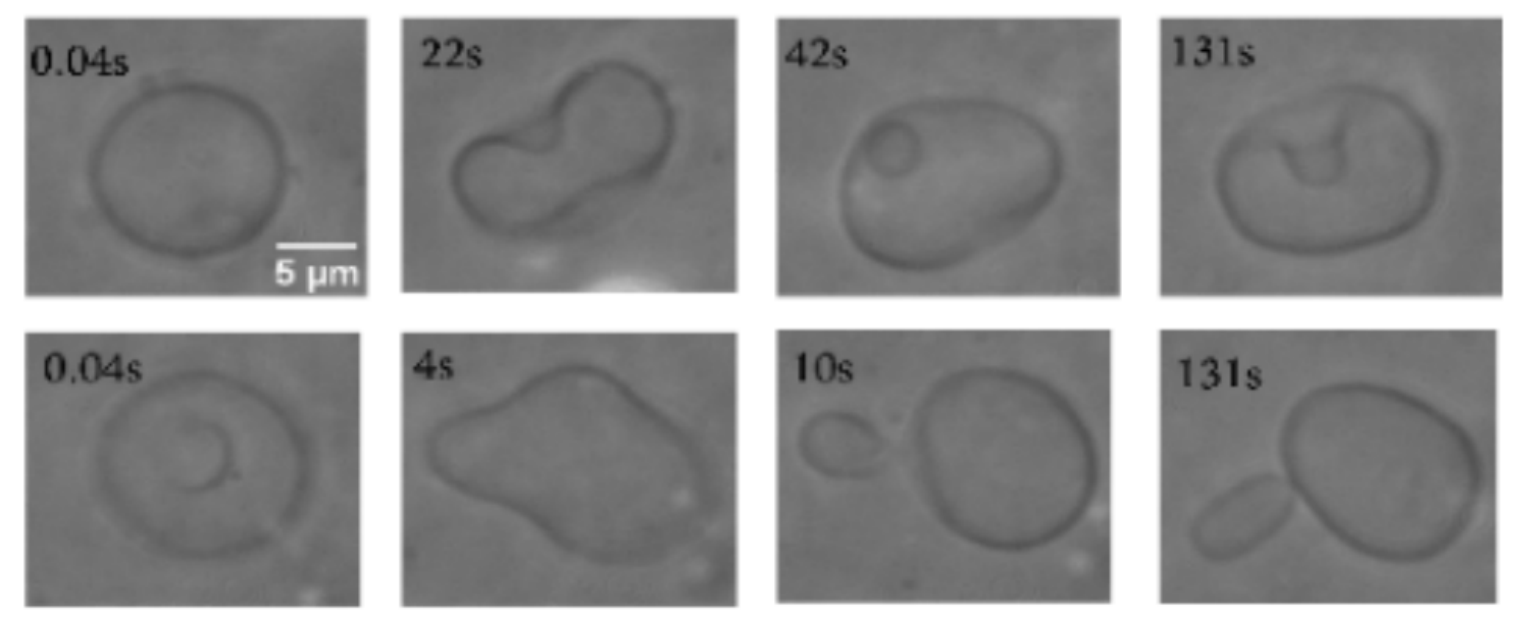}
 \caption{\textbf{Actuated membranes of polymersomes}: Snapshots of membrane shape transformations of a polymersome filled with 300 nm AuNP under laser illumination (top on, bottom off), showing strong morphological changes over time (unpublished data).}
 \label{fig:11}
 \end{figure}

We analyze the membrane dynamics of polymersomes under uniform heating. We assess changes in mechanical properties of the membrane using contour fluctuation spectroscopy to extract effective values of bending rigidity and effective membrane tension. To distinguish purely thermal effects from localized photothermal effects, we initially examined the response of polymersomes to heating using a Peltier-controlled stage. The results at increasing temperatures suggest that the low-mode region of the fluctuation spectrum is primarily affected, i.e., the effective membrane tension—while the bending rigidity remains of the same order of magnitude. When subjected to photothermal effects, AuNP-filled polymersomes similarly exhibit a change in mechanical properties, tunable by laser power, making the polymersomes appear mechanically stiffer at higher powers. Importantly, laser illumination of 300 nm AuNP-loaded polymersomes reveals complex and pathway-dependent morphological transformations. The polymersomes evolve from an initially quasi-spherical shape~(Figure~\ref{fig:11}) toward non-axisymmetric deformations, dumbbell-like intermediates, stomatocyte-like invaginations, and pearling instabilities containing the AuNPs. Upon switching the laser off, the process relaxes slowly to the initial state, or remains locked in another configuration, indicating that the membrane may access distinct morphological branches during relaxation, which may be related to the morphological diagram observed for polymersomes~\cite{rikken2016}.

\section{Engineering cell motility} ~\label{sec:7}

The membrane fluidity, composition, bending rigidity, and tension govern how cells deform, reorganize, and respond to internal and external forces, ultimately controlling the motion of the cell~\cite{gracia2010,steinkuhler2019}. We explore how motility emerges from activity upon external actuation of the membrane.

\subsection{Collective actuation of motors drives propulsion}

\begin{figure}[tbp]
 \includegraphics[width=\linewidth]{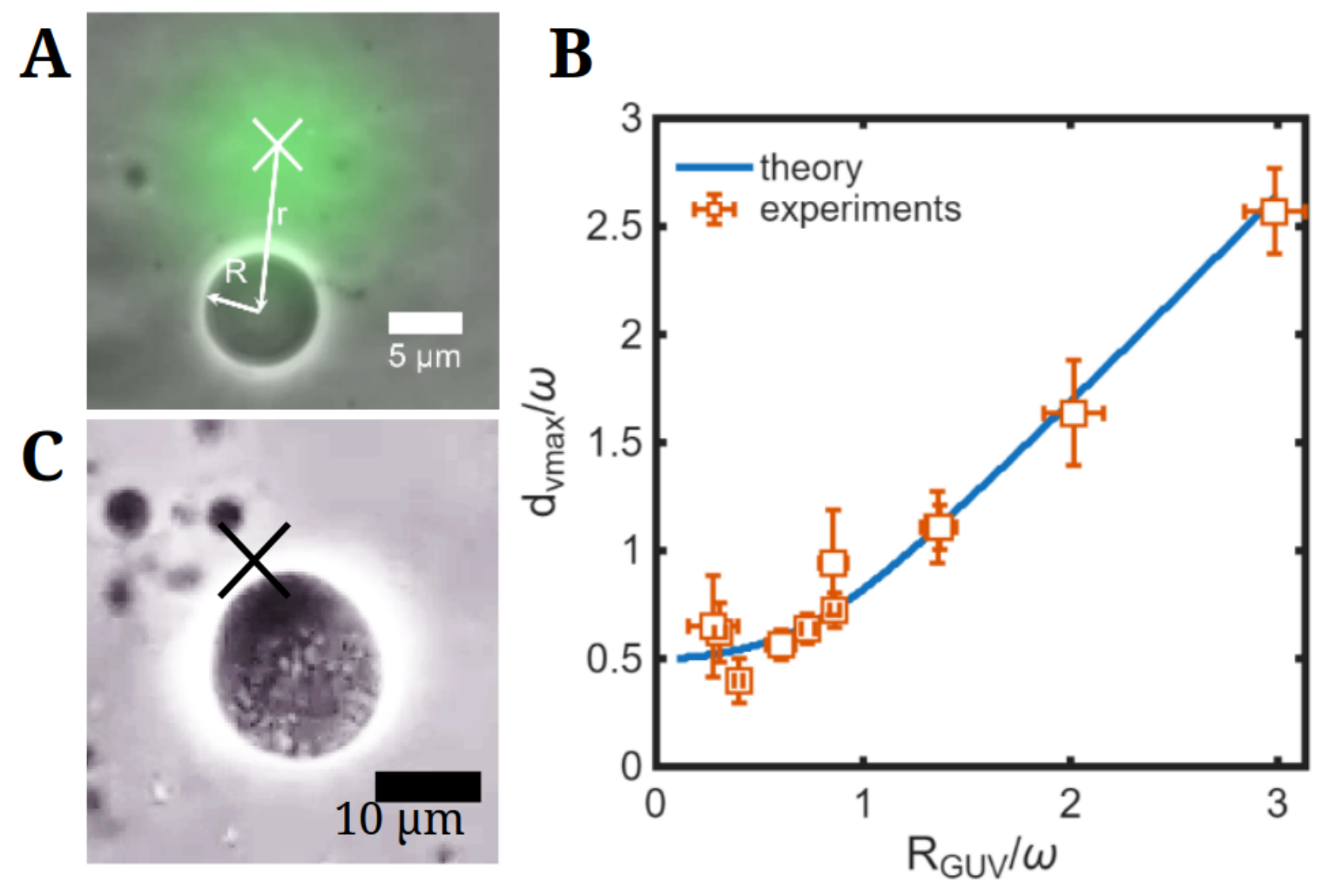}
 \caption{\textbf{Motor induced motility}: (A) Scheme of the process for analyzing the migration of a polymersome of radius $R$ at distance $r$ from the center of the laser spot, shown in green (B) Distance of the maximum velocity recorded during the migration against the radius of the empty polymersome, normalized by the laser waist, showing perfect agreement with the theory in the ray optics regime. (C) Snapshot showing the deformation of a polymersome and inhomogeneous distribution of gold nanoparticles inside, indicative of a coupling between internal organization and the dynamics of the polymersome. In A,C, the laser spot position is indicated by the cross (unpublished data)}
 \label{fig:12}
 \end{figure}

Interestingly, the nanomotor actuation mechanism presented in section~\ref{sec:6} impacts the propulsion of the polymersome. We measured the migration dynamics of polymersomes under non-uniform light, i.e., light gradients such as Gaussian beam illumination. Empty polymersomes are attracted towards the center of the laser spot, a behavior well-described by the classical theory of optical trapping, as initially developed by Ashkin in the ray optics regime~\cite{ashkin1997}, for all investigated radii. In contrast, polymersomes filled with AuNPs exhibit a non-trivial behavior. They are either attracted or repelled depending on the AuNP concentration, laser intensity, and reach velocities nearly an order of magnitude higher than empty polymersomes (Figure~\ref{fig:12}). These effects result from a strong nonlinear coupling between the internal organization of the nanoparticles, the resulting optothermal effects (tuning membrane properties and inducing thermophoretic migration) and the optical forces acting on both the AuNPs and the membrane. 

\subsection{Membrane phase separation to engineer motility}

We investigate how membrane organization and mechanics couple to externally supplied energy to control the motion of synthetic cells. Using giant unilamellar vesicles (GUVs) as a model system, we explore how controlled electric perturbations generate nonequilibrium behaviors mimicking the run and tumble motion observed in their living counterparts.

\begin{figure}[tbp]
 \includegraphics[width=\linewidth]{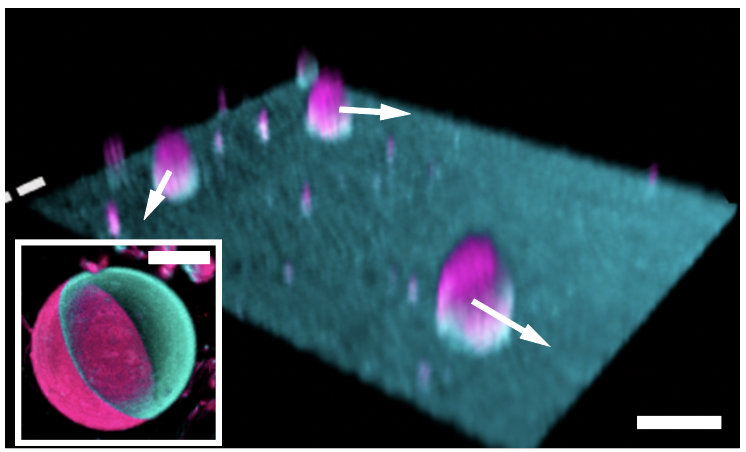}
 \caption{\textbf{Phase transition at the membrane promotes propulsion}: Janus lipid vesicles moving close to a substrate under the actuation of and AC electric field. The arrow depics the velocity vector, scale bar: 15 microns. Inset: confocal image of the Janus vesicle with the Lo and Ld lipid domains (scale-bar: 10 microns). Reprinted with permission from~\cite{willems2025}}
 \label{fig:7}
 \end{figure}

As a first step toward understanding how membrane properties regulate the dynamics of synthetic cells, we investigated the coupling between membrane fluidity, phase organization, and externally driven motility~\cite{willems2025}. To this end, we employed phase-separated giant unilamellar vesicles (GUVs), in which coexisting liquid-ordered (Lo) and liquid-disordered (Ld) domains introduce spatial variations in membrane properties. By applying external AC electric fields, we induced directed motion close to the surface and explored how the reorganization of lipid domains influences the resulting dynamics (Figure~\ref{fig:7}). Remarkably, these Janus GUVs exhibit run-and-tumble behavior, characterized by periods of persistent directional motion interrupted by abrupt reorientations. This behavior emerges from the coupling between field-induced propulsion and the dynamic rearrangement of membrane domains, demonstrating that membrane organization can directly regulate the motility of artificial cells. These findings establish membrane fluidity and phase behavior as key physical parameters for controlling nonequilibrium dynamics, providing a route toward synthetic cells whose motion is governed by their internal material properties.

Overall, these results demonstrate that engineering vesicles and polymersomes through membrane design or through energy-transducing nanomotors constitute a promising platform to design artificial cell-like systems with a tunable mechano-responsive behavior.

\section{Control of genome in compartments} ~\label{sec:8}

A key component of a living cell is its ability to self-replicate~\cite{Luisi2006} with a transfer of identity towards the offsprings. Genetic information, and the molecular machines required to express its encoded functions is key to this process in living systems as we know them. A clear objective of the bottom-up synthetic biology community is to design a cell bearing a precisely defined and tightly controlled blueprint. In this regard, the blueprint is genetic information, assimilated to a `program' that is `booted-up' inside of a machine, the cell compartment, to guarantee all its functions and its autonomy. The nature of that program remains to be determined, both in its content (the functions that the cell must carry out), but also in its form. When designing an entire genetic program from the ground up, what should the nature and layout of the coding and regulatory sequences be to reach the level of expression necessary to sustain cell viability? How can we fine-tune the interactions of those sequences with one another, as well as the interactions of their protein products with one another? Moreover, what can we expect of their interactions with the rest of the machine, and how do we ensure that they are recognizable and workable by the machine in the first place? Indeed, the matters of encoding and decoding genetic information go hand in hand.

Here, we explore the potential of genome transplantation. This technique is identified as the download of an entire genome in a cell in such a way that it replaces the endogenous genome to take control of the cell. Thus far, this phenomenon has only been demonstrated in the bacteria belonging to the Mollicutes class, which have acquired the title of `natural minimal cells' because of their apparent simplicity (small size, lack of a cell wall, minimalist metabolism, reduced genomes). They are the simplest and smallest known living forms capable of self-replication outside of a host, and an appropriate model to study minimal conditions for life. The experiments conducted on Mollicutes at the J. Craig Venter Institute in the 2010s resulted in the first living cell controlled by a synthetic genome, thanks to the use of genome transplantation~\cite{lartigue2007b, lartigue2009a, gibson2010e}. 

To perform genome transplantation, one isolates and transfers the genome of a given species, called the donor genome, into the cytoplasm of a recipient cell which can be of a different species. This process generates cells bearing the genotype and phenotype of the donor species and triggers a complete species change of the cell compartment. This method is revolutionary for the production of synthetic cells: it demonstrates the possibility of establishing a new program in a pre-existing machine, and delivers a framework to determine the molecular interactions that are essential for genetic information processing. Furthermore, the limitations of this phenomenon is a good place to start looking into how a program and the host machine interact. Indeed, genome transplantation is limited to a handful of Mollicutes donor genomes in the recipient cell \textit{Mycoplasma capricolum subsp. capricolum (Mcap)}; its expansion to other species is held back by some intrinsic rules that seem to govern it. Notably, it has been shown that genome transplantation efficiency decreases when the phylogenetic distance between donor genome and recipient cell increases~\cite{labroussaa2016a}, suggesting that such rules may be rooted in the notion of compatibility between donor and recipient species.

We hypothesized that the ability of the recipient cell to transcribe the donor genome is an early and essential factor in cell boot-up. To characterize the interactions between the transcription machineries of distinct species, we designed two complementary strategies involving two Mollicutes species that have previously been used for genome transplantation: \textit{Mesoplasma florum (M. florum)} as donor and \textit{Mcap} as recipient. First, we expressed five recombinant \textit{M. florum} RNA polymerase (RNAP) protein subunits in \textit{Mcap}, mimicking the unique cellular environment of transplantation. When the entire RNAP protein complexes of both species were expressed simultaneously, we observed mutations in the coding sequences of the catalytic subunits RpoBMfl and RpoCMfl of \textit{M. florum}, suggesting the presence of those foreign protein subunits is disadvantageous to \textit{Mcap}. In parallel, we aimed to evaluate the compatibility between a genome and the RNAP expressing it by replacing all the coding sequences of \textit{Mcap}'s RNAP by \textit{M. florum}'s. We discovered that \textit{Mcap} is functional when it expresses only the \textit{M. florum} RNAP, as well as several, but not all hybrid combinations of \textit{Mcap} and \textit{M. florum} RNAP subunits. In particular, we confirmed that the RpoB/RpoCMfl subunits are incompatible with the remaining \textit{Mcap} RNAP subunits. From those two independent approaches, we concluded that the formation of hybrid transcription machineries is toxic for the cell and decreases the probability of selecting a viable transplanted cell.

The lessons learned with this genome transplantation model allow us to establish parallels with the boot-up of an artificial genome in an artificial compartment, and deduce two complementary conclusions usable as guidelines for future designs: first, in a single compartment, cross-talk between proteins of different genetic origin may be a death sentence for the cell (as with \textit{Mcap} expressing both \textit{Mcap} and \textit{M. florum} RNAP subunits); second, molecular systems may be more flexible than we initially imagined with genetic components of a different genetic origin: for transcription at least, \textit{M. florum} RNAP is capable of transcribing the Mcap genome, while \textit{Mcap} RNAP is seemingly capable of transcribing the \textit{M. florum} donor genome as evidenced by the acquisition of transplants upon GT~\cite{JittaseviPHD}.




%


\section{Enabling technologies} \label{sec:9}

Three main enabling technologies are key in our approach for the development of synthetic cell research: microfluidics, 3d-bioprinting and cell free expression systems.

Microfluidics, and particularly droplet-based microfluidics, offers precise control over the generation, manipulation, and analysis of cell-like compartments at the microscale~\cite{moragues2023}. By encapsulating biological and biochemical components within highly uniform, monodisperse compartments, microfluidic systems provide well-defined microenvironments for organizing in space and time biochemical reactions and cellular functions under controlled conditions (see e.g. sections~\ref{sec:3},~\ref{sec:4},~\ref{sec:5}). Droplet-microfluidics acts as versatile assembly lines for the formation of biomolecular condensates, such as coacervates, through liquid–liquid phase separation, as well as for the assembly of membrane-bound vesicles, including liposomes and other vesicular structures that mimic key features of biological cells~\cite{beneyton2020b, weiss2018d}. The integration of droplet generation with downstream manipulation, reagent delivery, and compartment fusion enables the systematic reconstruction and study of increasingly complex synthetic cell systems, while facilitating high-throughput experimentation and the exploration of large parameter spaces. Furthermore, microfluidics offers powerful analytical capabilities through the integration of optical microscopy, fluorescence-based assays, spectroscopic techniques, and real-time monitoring, enabling the quantitative characterization of compartment formation, molecular partitioning, reaction kinetics, membrane properties, and functional outputs. Microfluidics provides means to manipulate vesicles and actuate dispersed systems using forcing by electric field, temperature, light\dots (see e.g. sections~\ref{sec:5},~\ref{sec:6},~\ref{sec:7}). 

Microfluidics can be viewed not only as an `assembly line' for synthetic cells, but also as a powerful platform for producing other advanced particles and for engineering the 2D and 3D architectures needed to compartmentalize, perfuse, stimulate, and analyze living systems, whether natural or synthetic. Digital microfluidics benefits materials science to generate particles (including emulsions, multiple emulsions, polymer and hydrogel microparticles, capsules, and encoded hydrogel particles) with controlled size and composition~\cite{shang2017,breukers2025}. When used alongside synthetic cells, simple liquid droplets can act as reservoirs or sinks for nutrients, metabolites, and other molecules. More importantly, the intrinsic hierarchy of multiple emulsions enables timed release or capture, as well as reaction cascades. Once converted into solid, porous, Janus, or other structured materials, microfluidic particles can serve several functions. Passive but functionalizable, they can capture molecules or metabolites for detection, or support extracellular catalysis and may have an important mechanical role: polymer particles with tunable stiffness can locally modify the rigidity or viscoelasticity of a matrix, thereby influencing the fate of synthetic cells. Hydrogel particles are particularly promising for engineering living systems ~\cite{abrishamkar2026}. Because hydrogels can mimic key properties of the extracellular matrix, they can host or support cell cultures, from simple 3D culture formats~\cite{lenzi2026} to highly patterned materials that control stiffness, functionalization, and other competing properties. Hydrogel capsules also contrast strongly with bi- or multiphasic droplets: their chemostat-like design supports both encapsulation and selective exchange with the environment~\cite{alessandri2013a}. Beyond particle fabrication, microfluidics is a powerful tool for biomimicry, particularly in organ-on-chip and organoid technologies~\footnote{Organ-on-chip platforms are designed to reproduce tissue-level functions by controlling cell–tissue interactions, mechanical cues, and perfusion.}. These systems are approaching the maturity needed to incorporate synthetic cells. Today, physiological microsystems integrate living components (cell lines, stem cells, primary cells, spheroids, and organoids) within functional microfluidic architectures that provide compartmentalization, vascularization, flow, and multiple stimuli, including pressure, temperature, and deformation~\cite{zhao2023}. Such engineering enables long-term 2D and 3D cell culture under finely controlled flow. It allows efficient management of hydrodynamic forces and of nutrient, gas, and metabolite transport. Static or dynamic constraints can also be applied locally, for example through temperature gradients, peristalsism, or anoxia, to increase physiological relevance. Sensors and actuators provide real-time readouts of system state and enable feedback control of the evolving microenvironment. 

At larger scales, bioprinting can structure bio-compatible materials to support the culture of selected cells, synthetic or natural, in volume or on surfaces. Programmable deposition can create gradients in mesh size, matrix composition, and cell density, as well as pseudo-vascularized regions, thereby imposing regional functions. The combination of microfluidics and bioprinting is therefore best understood as a hierarchical fabrication strategy: microfluidics provides dynamic physiology, flow, gradients, and shear stress, while bioprinting provides programmable anatomy. In terms of materials, elastin-like polypeptides, a class of biopolymers derived from the hydrophobic domain of elastin, have demonstrated exceptional biocompatibility and versatility for bioprinting applications~\cite{dai2021}. Formulations based on elastin-like polypeptides can further be used in regenerative medicine applications where synthetic cells will probably have a functional role in the future, paving the way for advanced regenerative therapies. 

The goal is no longer simply to create an isolated synthetic cell, but to design a microtissue containing spatially organized producer and consumer zones, exchange interfaces, perfusion channels, mechanical supports, and regulatory feedback loops. This approach could enable a progressive transition from isolated synthetic cells to functional microtissues and then to perfused proto-organs. These would not be exact replicas of organs, but programmable, reproducible, and controllable biohybrid systems with multiple cell populations, vascular or perfused equivalents, tissue interfaces, mechanical responses, and distributed regulation.

On the biochemical side, cell-free expression enables the \textit{in vitro} bio-synthesis of proteins and other biomolecules without the need for intact, living cells~\cite{hunt2025}. By harnessing the transcriptional and translational machinery of cellular extracts or reconstituted biochemical systems, cell-free expression platforms provide a flexible and accessible means of implementing gene expression within controlled, cell-like environments. These systems typically combine DNA or RNA templates with ribosomes, enzymes, nucleotides, amino acids, energy-regeneration substrates, and other essential molecular components to support protein synthesis outside the cellular context. A major advantage of cell-free expression is the ability to directly manipulate the biochemical composition and physicochemical conditions of the reaction environment, facilitating the systematic investigation of gene expression, metabolic pathways, and regulatory networks. In synthetic cell research, cell-free expression can be integrated into a variety of compartmentalized architectures, including microfluidic droplets, coacervates, and membrane-bound vesicles, providing a means to couple genetic information with compartmentalized biochemical activity~\cite{Sierra2026, meier2026}. This integration enables the development of increasingly complex synthetic systems capable of expressing fluorescent reporters, producing functional proteins, and implementing rudimentary cellular functions, especially considering the potential offered by Genome Transplantation techniques (see e.g. sections~\ref{sec:8}). Furthermore, cell-free expression offers powerful analytical capabilities through real-time monitoring of gene expression, quantitative fluorescence measurements, kinetic analysis, and high-throughput screening of reaction conditions and genetic constructs. By bypassing the constraints associated with cellular viability, membrane transport, and endogenous regulatory mechanisms, cell-free expression provides a highly tunable platform for dissecting and reconstructing fundamental biological processes. 

These three enabling technologies are highly complementary for the assembly of single-cell mimics (microfluidics), with a high level of programmability (cell-free expression, genome transplant) and for their integration in complex manufacturable environments (3d printing). Their combined use will provide in the future new strategies for synthetic cell research.





\section{Conclusions}

The summary presented here is representative of the current concerted and original effort of the local community represented in the Research Network Frontiers of Life, with an emphasis on the work of the doctoral students involved in the network over the funding period~\cite{JittaseviPHD,lescouarnecPHD,beauperePHD,peyraud-vicrePHD,debPHD,chaudhuriPHD}. 

We have tackled research activities in the context of the assembly of synthetic cells. We focused on several key fundamental processes that contribute to the functions and properties of living systems, compartmentalization, energy management, communication through molecular transport and genetic information encoding. We highlight specifically the crucial role of phase separation in providing regulation mechanisms and switchable systems as well as the key role of out-of-equilibrium conditions to properly reproduce the behaviours of living systems in model systems. In the long term, the ability to control and engineer these out-of-equilibrium conditions is expected to provide a conceptual framework complementary to the structural, biochemical and genetic engineering required for the assembly of a life-like cell. 

The assembly of a synthetic cell is still ahead and requires a large scale organisation and an interdisciplinary approach. Our community will in the future continue to contribute in collaboration with other networks of researchers (MaxSynBio, SynCell-EU, SynCell-FR, FabriCell, Build-a-Cell\dots) to this grand challenge of the XXI\up{st} Century - the construction of synthetic cells - which, we believe, will result in new biotechnologies in the future. 


\printCOI
\printESM


\bibliographystyle{crchim}
 \bibliography{FrontiersOfLife.bib}

\def\bysame{\leavevmode ---------\thinspace}
\makeatletter\if@francais\providecommand{\og}{<<~}\providecommand{\fg}{~>>}
\else\gdef\og{``}\gdef\fg{''}\fi\makeatother
\def\cdrandname{\&}
\providecommand\cdrnumero{no.~}
\providecommand{\cdredsname}{eds.}
\providecommand{\cdredname}{ed.}
\providecommand{\cdrchapname}{chap.}
\providecommand{\cdrmastersthesisname}{Memoir}
\providecommand{\cdrphdthesisname}{PhD Thesis}
\begin{thebibliography}{100}

\bibitem{catlingHowEarthsAtmosphere2005}
D.~C. Catling, M.~W. Claire, {\og How {{Earth}}'s Atmosphere Evolved to an Oxic
  State: {{A}} Status Report\fg}, \emph{Earth and Planetary Science Letters},
  2005,  \textbf{237}, \cdrnumero 1, 1-20.

\bibitem{Cleland2002}
C.~Cleland, C.~Chyba, {\og Defining {{Life}}\fg}, \emph{Origins of Life and
  Evolution of the Biosphere}, 2002,  \textbf{32}, 387-393.

\bibitem{bartlettDefiningLyfeUniverse2020}
S.~Bartlett, M.~L. Wong, {\og Defining Lyfe in the Universe: {{From}} Three
  Privileged Functions to Four Pillars\fg}, \emph{Life}, 2020,  \textbf{10},
  \cdrnumero 4.

\bibitem{schrodingerWhatLife1967}
E.~Schrodinger, \emph{What Is Life?}, Cambridge University Press, Cambridge,
  1967.

\bibitem{macheryWhyStoppedWorrying2012}
E.~Machery, {\og Why {{I}} Stopped Worrying about the Definition of Life... and
  Why You Should as Well\fg}, \emph{Synthese}, 2012,  \textbf{185}, \cdrnumero
  1, 145-164.

\bibitem{schultzRudolfVirchow2008}
M.~Schultz, {\og Rudolf {{Virchow}}\fg}, \emph{Emerging Infectious Diseases},
  2008,  \textbf{14}, \cdrnumero 9, 1480-1481.

\bibitem{gibson2010e}
D.~G. Gibson, J.~I. Glass, C.~Lartigue, V.~N. Noskov, R.-Y. Chuang, M.~A.
  Algire, G.~A. Benders, M.~G. Montague, L.~Ma, M.~M. Moodie, C.~Merryman,
  S.~Vashee, R.~Krishnakumar, N.~{Assad-Garcia}, C.~{Andrews-Pfannkoch}, E.~A.
  Denisova, L.~Young, Z.-Q. Qi, T.~H. {Segall-Shapiro}, C.~H. Calvey, P.~P.
  Parmar, C.~A. Hutchison, H.~O. Smith, J.~C. Venter, {\og Creation of a
  Bacterial Cell Controlled by a Chemically Synthesized Genome\fg},
  \emph{Science}, 2010,  \textbf{329}, \cdrnumero 5987, 52-56.

\bibitem{labroussaaTransplantationGenomesRedonner2019}
F.~Labroussaa, V.~Baby, S.~Rodrigue, C.~Lartigue, {\og {La transplantation de
  g\'enomes - Redonner vie \`a des g\'enomes bact\'eriens naturels ou
  synth\'etiques}\fg}, \emph{m\'edecine/sciences}, 2019,  \textbf{35},
  \cdrnumero 10, 761-770.

\bibitem{lartigueCreatingBacterialStrains2009}
C.~Lartigue, S.~Vashee, M.~A. Algire, R.-Y. Chuang, G.~A. Benders, L.~Ma, V.~N.
  Noskov, E.~A. Denisova, D.~G. Gibson, N.~{Assad-Garcia}, N.~Alperovich, D.~W.
  Thomas, C.~Merryman, C.~A. Hutchison, H.~O. Smith, J.~C. Venter, J.~I. Glass,
  {\og Creating Bacterial Strains from Genomes That Have Been Cloned and
  Engineered in Yeast\fg}, \emph{Science}, 2009,  \textbf{325}, \cdrnumero
  5948, 1693-1696.

\bibitem{lartigue2007b}
C.~Lartigue, J.~I. Glass, N.~Alperovich, R.~Pieper, P.~P. Parmar, C.~A.
  Hutchison, H.~O. Smith, J.~C. Venter, {\og Genome Transplantation in
  Bacteria: Changing One Species to Another\fg}, \emph{Science}, 2007,
  \textbf{317}, \cdrnumero 5838, 632-638.

\bibitem{clementStephaneLeducPhysique2018}
R.~Cl{\'e}ment, {\og {St\'ephane Leduc et la physique du vivant}\fg},
  \emph{Reflets de la physique}, 2018, \cdrnumero 58, 26-29.

\bibitem{oparinOriginLifeOrigin1965}
A.~I. Oparin, {\og The Origin of Life and the Origin of Enzymes\fg},
  \emph{Advances in Enzymology and Related Areas of Molecular Biology}, 1965,
  \textbf{27}, 347-380.

\bibitem{szostakSynthesizingLife2001}
J.~W. Szostak, D.~P. Bartel, P.~L. Luisi, {\og Synthesizing Life\fg}, 2001,
  387-390~pages.

\bibitem{buddinghArtificialCellsSynthetic2017b}
B.~C. Buddingh', J.~C.~M. {van Hest}, {\og Artificial {{Cells}}: {{Synthetic
  Compartments}} with {{Life}}-like {{Functionality}} and {{Adaptivity}}\fg},
  \emph{Accounts of Chemical Research}, 2017,  \textbf{50}, \cdrnumero 4,
  769-777.

\bibitem{bar-onBiomassDistributionEarth2018}
Y.~M. {Bar-On}, R.~Phillips, R.~Milo, {\og The Biomass Distribution on
  {{Earth}}\fg}, \emph{Proceedings of the National Academy of Sciences}, 2018,
  \textbf{115}, \cdrnumero 25, 6506-6511.

\bibitem{meng2020a}
F.~Meng, T.~Ellis, {\og The Second Decade of Synthetic Biology: 2010--2020\fg},
  \emph{Nature Communications}, 2020,  \textbf{11}, \cdrnumero 1, 5174.

\bibitem{kitanoSyntheticBiologyLearning2023}
S.~Kitano, C.~Lin, J.~L. Foo, M.~W. Chang, {\og Synthetic Biology: {{Learning}}
  the Way toward High-Precision Biological Design\fg}, \emph{PLOS Biology},
  2023,  \textbf{21}, \cdrnumero 4, e3002116.

\bibitem{luisiEmergenceLifeChemical2006}
P.~L. Luisi, \emph{The {{Emergence}} of {{Life}}: {{From Chemical Origins}} to
  {{Synthetic Biology}}}, Cambridge University Press, Cambridge, 2006.

\bibitem{schwilleMaxSynBioAvenuesCreating2018}
P.~Schwille, J.~Spatz, K.~Landfester, E.~Bodenschatz, S.~Herminghaus,
  V.~Sourjik, T.~Erb, P.~Bastiaens, R.~Lipowsky, A.~Hyman, P.~Dabrock, J.-C.
  Baret, T.~{Vidakovic-Koch}, P.~Bieling, R.~Dimova, H.~Mutschler, T.~Robinson,
  D.~Tang, S.~Wegner, K.~Sundmacher, {\og {{MaxSynBio}} - {{Avenues}} towards
  Creating Cells from the Bottom Up\fg}, \emph{Angewandte Chemie International
  Edition}, 2018.

\bibitem{lin2023c}
Z.~Lin, T.~Beneyton, J.-C. Baret, N.~Martin, {\og Coacervate {{Droplets}} for
  {{Synthetic Cells}}\fg}, \emph{Small Methods}, 2023,  \textbf{7}, \cdrnumero
  12, 2300496.

\bibitem{noireauxVesicleBioreactorStep2004b}
V.~Noireaux, A.~Libchaber, {\og A Vesicle Bioreactor as a Step toward an
  Artificial Cell Assembly\fg}, \emph{Proceedings of the National Academy of
  Sciences of the United States of America}, 2004,  \textbf{101}, \cdrnumero
  51, 17669-17674.

\bibitem{palivan2024}
C.~G. Palivan, L.~Heuberger, J.~Gaitzsch, B.~Voit, D.~Appelhans,
  B.~Borges~Fernandes, G.~Battaglia, J.~Du, L.~Abdelmohsen, J.~C.~M. Van~Hest,
  J.~Hu, S.~Liu, Z.~Zhong, H.~Sun, A.~Mutschler, S.~Lecommandoux, {\og
  Advancing {{Artificial Cells}} with {{Functional Compartmentalized Polymeric
  Systems}} - {{In Honor}} of {{Wolfgang Meier}}\fg}, \emph{Biomacromolecules},
  2024,  \textbf{25}, \cdrnumero 9, 5454-5467.

\bibitem{schvartzman2024}
C.~Schvartzman, E.~Ibarboure, A.~Martin, E.~Garanger, A.~Mutschler,
  S.~Lecommandoux, {\og Protocells {{Featuring Membrane-Bound}} and {{Dynamic
  Membraneless Organelles}}\fg}, \emph{Biomacromolecules}, 2024,  \textbf{25},
  \cdrnumero 7, 4087-4094.

\bibitem{banani2017a}
S.~F. Banani, H.~O. Lee, A.~A. Hyman, M.~K. Rosen, {\og Biomolecular
  Condensates: Organizers of Cellular Biochemistry\fg}, \emph{Nature Reviews
  Molecular Cell Biology}, 2017,  \textbf{18}, \cdrnumero 5, 285-298.

\bibitem{alberti2021}
S.~Alberti, A.~A. Hyman, {\og Biomolecular Condensates at the Nexus of Cellular
  Stress, Protein Aggregation Disease and Ageing\fg}, \emph{Nature Reviews
  Molecular Cell Biology}, 2021,  \textbf{22}, \cdrnumero 3, 196-213.

\bibitem{gouveia2022}
B.~Gouveia, Y.~Kim, J.~W. Shaevitz, S.~Petry, H.~A. Stone, C.~P. Brangwynne,
  {\og Capillary Forces Generated by Biomolecular Condensates\fg},
  \emph{Nature}, 2022,  \textbf{609}, \cdrnumero 7926, 255-264.

\bibitem{peyraud-vicre2025}
K.~{Peyraud-Vicr{\'e}}, C.~Dechamps, N.~Martin, V.~Desvergnes, {\og Coacervate
  {{Droplets Drive Organocatalyzed Aqueous C}}--{{C Bond Formation}} via
  {{Interfacial Activation}}\fg}, \emph{Journal of the American Chemical
  Society}, 2025,  \textbf{147}, \cdrnumero 41, 37337-37346.

\bibitem{dherbassy2024}
Q.~Dherbassy, R.~J. Mayer, J.~Moran, {\og Coenzymes in a Pre-Enzymatic
  Metabolism\fg}, \emph{Science Advances}, 2024,  \textbf{10}, \cdrnumero 38,
  eadr5357.

\bibitem{Kern1997}
D.~Kern, G.~Kern, H.~Neef, K.~Tittmann, M.~{Killenberg-Jabs}, C.~Wikner,
  G.~Schneider, G.~H{\"u}bner, {\og How {{Thiamine Diphosphate Is Activated}}
  in {{Enzymes}}\fg}, \emph{Science}, 1997,  \textbf{275}, \cdrnumero 5296,
  67-70.

\bibitem{debiais2020}
M.~Debiais, A.~Hamoud, R.~Drain, P.~Barth{\'e}l{\'e}my, V.~Desvergnes, {\og
  Bio-Inspired {{NHC}}-Organocatalyzed {{Stetter}} Reaction in Aqueous
  Conditions\fg}, \emph{RSC Advances}, 2020,  \textbf{10}, \cdrnumero 67,
  40709-40718.

\bibitem{peyraud-vicrePHD}
K.~{Peyraud-Vicr{\'e}}, {\og Catalyse Bio-Inspir\'ee En Gouttelettes Aqueuses :
  Vers Une Nouvelle Chimie Du Vivant\fg}, These de doctorat, Bordeaux, 2025.

\bibitem{swain2024}
B.~C. Swain, P.~Sarkis, V.~Ung, S.~Rousseau, L.~Fernandez, A.~Meltonyan, V.~E.
  Aho, D.~Mercadante, C.~D. Mackereth, M.~Aznauryan, {\og Disordered Regions of
  Human {{eIF4B}} Orchestrate a Dynamic Self-Association Landscape\fg},
  \emph{Nature Communications}, 2024,  \textbf{15}, \cdrnumero 1, 8766.

\bibitem{kar2022}
M.~Kar, F.~Dar, T.~J. Welsh, L.~T. Vogel, R.~K{\"u}hnemuth, A.~Majumdar,
  G.~Krainer, T.~M. Franzmann, S.~Alberti, C.~A.~M. Seidel, T.~P.~J. Knowles,
  A.~A. Hyman, R.~V. Pappu, {\og Phase-Separating {{RNA}}-Binding Proteins Form
  Heterogeneous Distributions of Clusters in Subsaturated Solutions\fg},
  \emph{Proceedings of the National Academy of Sciences}, 2022,  \textbf{119},
  \cdrnumero 28, e2202222119.

\bibitem{chaudhuriPHD}
P.~Chaudhuri, {\og D\'eveloppement d'un Mod\`ele Multidisciplinaire de
  S\'eparation de Phase Des Prot\'eines in Vitro et in Cellullo\fg}, These en
  pr\'eparation, Bordeaux, 2023.

\bibitem{lin2026}
Z.~Lin, T.~Beneyton, S.~Lafon, E.~R.~J. Granda, L.~Tian, A.~Baron, J.-C. Baret,
  N.~Martin, {\og Light Selects Non-Equilibrium States of Phase-Separated
  Droplets\fg}.

\bibitem{lopez-otin2013}
C.~{L{\'o}pez-Ot{\'i}n}, M.~A. Blasco, L.~Partridge, M.~Serrano, G.~Kroemer,
  {\og The {{Hallmarks}} of {{Aging}}\fg}, \emph{Cell}, 2013,  \textbf{153},
  \cdrnumero 6, 1194-1217.

\bibitem{joyner2016}
R.~P. Joyner, J.~H. Tang, J.~Helenius, E.~Dultz, C.~Brune, L.~J. Holt, S.~Huet,
  D.~J. M{\"u}ller, K.~Weis, {\og A Glucose-Starvation Response Regulates the
  Diffusion of Macromolecules\fg}, \emph{eLife}, 2016,  \textbf{5}, e09376.

\bibitem{munder2016}
M.~C. Munder, D.~Midtvedt, T.~Franzmann, E.~N{\"u}ske, O.~Otto, M.~Herbig,
  E.~Ulbricht, P.~M{\"u}ller, A.~Taubenberger, S.~Maharana, L.~Malinovska,
  D.~Richter, J.~Guck, V.~Zaburdaev, S.~Alberti, {\og A {{pH}}-Driven
  Transition of the Cytoplasm from a Fluid- to a Solid-like State Promotes
  Entry into Dormancy\fg}, \emph{eLife}, 2016,  \textbf{5}, e09347.

\bibitem{delarue2018}
M.~Delarue, G.~P. Brittingham, S.~Pfeffer, I.~V. Surovtsev, S.~Pinglay, K.~J.
  Kennedy, M.~Schaffer, J.~I. Gutierrez, D.~Sang, G.~Poterewicz, J.~K. Chung,
  J.~M. Plitzko, J.~T. Groves, C.~{Jacobs-Wagner}, B.~D. Engel, L.~J. Holt,
  {\og {{mTORC1 Controls Phase Separation}} and the {{Biophysical Properties}}
  of the {{Cytoplasm}} by {{Tuning Crowding}}\fg}, \emph{Cell}, 2018,
  \textbf{174}, \cdrnumero 2, 338-349.e20.

\bibitem{romila2021}
C.~A. Romila, S.~Townsend, M.~Malecki, S.~Kamrad, M.~{Rodr{\'i}guez-L{\'o}pez},
  O.~Hillson, C.~Cotobal, M.~Ralser, J.~B{\"a}hler, {\og Barcode Sequencing and
  a High-Throughput Assay for Chronological Lifespan Uncover Ageing-Associated
  Genes in Fission Yeast\fg}, \emph{Microbial Cell}, 2021,  \textbf{8},
  \cdrnumero 7, 146-160.

\bibitem{sawyer2021}
D.~P. Sawyer, A.~{Bar-Zion}, A.~Farhadi, S.~Shivaei, B.~Ling,
  A.~{Lee-Gosselin}, M.~G. Shapiro, {\og Ultrasensitive Ultrasound Imaging of
  Gene Expression with Signal Unmixing\fg}, \emph{Nature Methods}, 2021,
  \textbf{18}, \cdrnumero 8, 945-952.

\bibitem{hondele2020}
M.~Hondele, S.~Heinrich, P.~De~Los~Rios, K.~Weis, {\og Membraneless Organelles:
  Phasing out of Equilibrium\fg}, \emph{Emerging Topics in Life Sciences},
  2020,  \textbf{4}, \cdrnumero 3, 331-342.

\bibitem{alberti2019}
S.~Alberti, D.~Dormann, {\og Liquid-{{Liquid Phase Separation}} in
  {{Disease}}\fg}, \emph{Annual Review of Genetics}, 2019,  \textbf{53},
  171-194.

\bibitem{schopper2017}
S.~Schopper, A.~Kahraman, P.~Leuenberger, Y.~Feng, I.~Piazza, O.~M{\"u}ller,
  P.~J. Boersema, P.~Picotti, {\og Measuring Protein Structural Changes on a
  Proteome-Wide Scale Using Limited Proteolysis-Coupled Mass Spectrometry\fg},
  \emph{Nature Protocols}, 2017,  \textbf{12}, \cdrnumero 11, 2391-2410.

\bibitem{hofmann2020}
K.~Hofmann, {\og The {{Evolutionary Origins}} of {{Programmed Cell Death
  Signaling}}\fg}, \emph{Cold Spring Harbor Perspectives in Biology}, 2020,
  \textbf{12}, \cdrnumero 9, a036442.

\bibitem{lopatina2020}
A.~Lopatina, N.~Tal, R.~Sorek, {\og Abortive {{Infection}}: {{Bacterial
  Suicide}} as an {{Antiviral Immune Strategy}}\fg}, \emph{Annual Review of
  Virology}, 2020,  \textbf{7}, \cdrnumero 1, 371-384.

\bibitem{maekawa2023}
T.~Maekawa, H.~Kashkar, N.~S. Coll, {\og Dying in Self-Defence: A Comparative
  Overview of Immunogenic Cell Death Signalling in Animals and Plants\fg},
  \emph{Cell Death \& Differentiation}, 2023,  \textbf{30}, \cdrnumero 2,
  258-268.

\bibitem{saupe2020}
S.~J. Saupe, {\og Amyloid {{Signaling}} in {{Filamentous Fungi}} and
  {{Bacteria}}\fg}, \emph{Annual Review of Microbiology}, 2020,  \textbf{74},
  \cdrnumero 1, 673-691.

\bibitem{ibarlosa2025}
L.~Ibarlosa, S.~Dheur, C.~Sanchez, S.~Deb, A.~{Granger-Farbos}, V.~Coustou,
  M.~Berbon, A.~{Massoni-Laporte}, B.~Salin, C.~Blancard, C.~Saragaglia, X.~M.
  Tene, N.~Dufour, O.~Rossier, B.~Habenstein, L.~Debarbieux, B.~Kauffmann,
  A.~Loquet, S.~J. Saupe, {\og Characterization of an Amyloid-Based Antiphage
  Defence System in {{Escherichia}} Coli\fg}, \emph{Nature Microbiology}, 2025,
   \textbf{10}, \cdrnumero 9, 2168-2178.

\bibitem{debPHD}
S.~Deb, {\og Reconstitution et \'Etudes Structurelles et Fonctionelles Des
  Signalosomes Impliqu\'es Dans La R\'egulation de La Mort Cellulaire
  Encapsul\'es Dans Des Protocellules\fg}, These en pr\'eparation, Bordeaux,
  2023.

\bibitem{audeoud2026}
{Audeoud et al.}, {\og Microfluidic Insights into the Fibrillation of
  Amyloids\fg}, \emph{ACS omega}, 2026,  \textbf{in press}.

\bibitem{walsh2018}
C.~T. Walsh, B.~P. Tu, Y.~Tang, {\og Eight {{Kinetically Stable}} but
  {{Thermodynamically Activated Molecules}} That {{Power Cell Metabolism}}\fg},
  \emph{Chemical Reviews}, 2018,  \textbf{118}, \cdrnumero 4, 1460-1494.

\bibitem{hwang2024}
S.-W. Hwang, M.~Kim, A.~P. Liu, {\og Towards {{Synthetic Cells}} with
  {{Self}}-{{Producing Energy}}\fg}, \emph{ChemPlusChem}, 2024,  \textbf{89},
  \cdrnumero 10, e202400138.

\bibitem{xu2022a}
C.~Xu, N.~Martin, M.~Li, S.~Mann, {\og Living Material Assembly of
  Bacteriogenic Protocells\fg}, \emph{Nature}, 2022,  \textbf{609}, \cdrnumero
  7929, 1029-1037.

\bibitem{xu2025}
C.~Xu, M.~Li, N.~Martin, S.~Mann, {\og Construction of Complex Bacteriogenic
  Protocells from Living Material Assembly\fg}, \emph{Nature Protocols}, 2025,
  \textbf{20}, \cdrnumero 9, 2586-2617.

\bibitem{beneyton2018c}
T.~Beneyton, D.~Krafft, C.~Bednarz, C.~Kleineberg, C.~Woelfer, I.~Ivanov,
  T.~{Vidakovi{\'c}-Koch}, K.~Sundmacher, J.-C. Baret, {\og Out-of-Equilibrium
  Microcompartments for the Bottom-up Integration of Metabolic Functions\fg},
  \emph{Nature Communications}, 2018,  \textbf{9}, \cdrnumero 1, 2391.

\bibitem{Bioenergetics22014}
G.~D. Nicholls, J.~S. Ferguson, \emph{Bioenergetics 2}, 2014.

\bibitem{li2022}
C.~Li, X.~Zhang, B.~Yang, F.~Wei, Y.~Ren, W.~Mu, X.~Han, {\og Reversible
  {{Deformation}} of {{Artificial Cell Colonies Triggered}} by {{Actin
  Polymerization}} for {{Muscle Behavior Mimicry}}\fg}, \emph{Advanced
  Materials}, 2022,  \textbf{34}, \cdrnumero 34, 2204039.

\bibitem{nair2023}
K.~S. Nair, H.~Bajaj, {\og Advances in Giant Unilamellar Vesicle Preparation
  Techniques and Applications\fg}, \emph{Advances in Colloid and Interface
  Science}, 2023,  \textbf{318}, 102935.

\bibitem{weissig2021}
V.~Weissig, M.~Edeas (\cdredsname), \emph{Mitochondrial {{Medicine}}:
  {{Volume}} 2: {{Assessing Mitochondria}}}, Methods in {{Molecular Biology}},
  vol. 2276, Springer US, New York, NY, 2021.

\bibitem{lefrancois2021}
P.~Lefran{\c c}ois, J.~Santolini, S.~Arbault, {\og Electroanalysis at a
  {{Single Giant Vesicle Generating Enzymatically}} a {{Reactive Oxygen
  Species}}\fg}, \emph{Analytical Chemistry}, 2021,  \textbf{93}, \cdrnumero
  39, 13143-13151.

\bibitem{beauperePHD}
{\og D\'eveloppement de Cellules Artificielles Bio\'energ\'etiques \`a Base de
  Mitochondries Encapsul\'ees Dans Des Microcompartiments \textbar{}
  {{Theses}}.Fr\fg}, https://theses.fr/2025BORD0457.

\bibitem{miller2020a}
T.~E. Miller, T.~Beneyton, T.~Schwander, C.~Diehl, M.~Girault, R.~McLean,
  T.~Chotel, P.~Claus, N.~S. Cortina, J.-C. Baret, T.~J. Erb, {\og
  Light-Powered {{CO}}{\textsubscript{2}} Fixation in a Chloroplast Mimic with
  Natural and Synthetic Parts\fg}, \emph{Science}, 2020,  \textbf{368},
  \cdrnumero 6491, 649-654.

\bibitem{schwander2016c}
T.~Schwander, L.~Schada Von~Borzyskowski, S.~Burgener, N.~S. Cortina, T.~J.
  Erb, {\og A Synthetic Pathway for the Fixation of Carbon Dioxide in
  Vitro\fg}, \emph{Science}, 2016,  \textbf{354}, \cdrnumero 6314, 900-904.

\bibitem{alexandre2023}
A.~Alexandre, M.~Lavaud, N.~Fares, E.~Millan, Y.~Louyer, T.~Salez,
  Y.~Amarouchene, T.~Gu{\'e}rin, D.~S. Dean, {\og Non-{{Gaussian Diffusion Near
  Surfaces}}\fg}, \emph{Physical Review Letters}, 2023,  \textbf{130},
  \cdrnumero 7, 077101.

\bibitem{detournemire2026}
G.~De~Tournemire, N.~Fares, Y.~Amarouchene, T.~Salez, {\og First-Passage
  Statistics of Confined Colloids\fg}, \emph{Physical Review Research}, 2026,
  \textbf{8}, \cdrnumero 3, 033309.

\bibitem{fares2024}
N.~Fares, M.~Lavaud, Z.~Zhang, A.~Jha, Y.~Amarouchene, T.~Salez, {\og
  Observation of {{Brownian}} Elastohydrodynamic Forces Acting on Confined Soft
  Colloids\fg}, \emph{Proceedings of the National Academy of Sciences}, 2024,
  \textbf{121}, \cdrnumero 42, e2411956121.

\bibitem{lescouarnec2025}
R.~Le~Scouarnec, E.~Ibarboure, L.~Alembik, S.~Lecommandoux, J.~Leblond~Chain,
  C.~Bonduelle, {\og Membrane-{{Anchored Polyproline Provides Controlled
  Microdomain Formation}} and {{Permeability}} in {{Lipid Vesicles}}\fg},
  \emph{Journal of the American Chemical Society}, 2025,  \textbf{147},
  \cdrnumero 28, 24213-24219.

\bibitem{lu2022a}
Y.~Lu, G.~Allegri, J.~Huskens, {\og Vesicle-Based Artificial Cells: Materials,
  Construction Methods and Applications\fg}, \emph{Materials Horizons}, 2022,
  \textbf{9}, \cdrnumero 3, 892-907.

\bibitem{Zhao2022}
J.~Zhao, Y.~Zhang, X.~Zhang, C.~Li, H.~Du, S.~M. S{\o}nderskov, W.~Mu, M.~Dong,
  X.~Han, {\og Mimicking {{Cellular Metabolism}} in {{Artificial Cells}}:
  {{Universal Molecule Transport}} across the {{Membrane}} through {{Vesicle
  Fusion}}\fg}, \emph{Analytical Chemistry}, 2022,  \textbf{94}, \cdrnumero 9,
  3811-3818.

\bibitem{shin2022}
J.~Shin, B.~D. Cole, T.~Shan, Y.~Jang, {\og Heterogeneous {{Synthetic
  Vesicles}} toward {{Artificial Cells}}: {{Engineering Structure}} and
  {{Composition}} of {{Membranes}} for {{Multimodal Functionalities}}\fg},
  \emph{Biomacromolecules}, 2022,  \textbf{23}, \cdrnumero 4, 1505-1518.

\bibitem{marguet2013d}
M.~Marguet, C.~Bonduelle, S.~Lecommandoux, {\og Multicompartmentalized
  Polymeric Systems: Towards Biomimetic Cellular Structure and Function\fg},
  \emph{Chemical Society Reviews}, 2013,  \textbf{42}, \cdrnumero 2, 512-529.

\bibitem{Huang2013}
J.~Huang, A.~Heise, {\og Stimuli Responsive Synthetic Polypeptides Derived from
  {{N}}-Carboxyanhydride ({{NCA}}) Polymerisation\fg}, \emph{Chemical Society
  Reviews}, 2013,  \textbf{42}, \cdrnumero 17, 7373-7390.

\bibitem{izgu2016}
E.~C. Izgu, A.~Bj{\"o}rkbom, N.~P. Kamat, V.~S. Lelyveld, W.~Zhang, T.~Z. Jia,
  J.~W. Szostak, {\og N-{{Carboxyanhydride-Mediated Fatty Acylation}} of
  {{Amino Acids}} and {{Peptides}} for {{Functionalization}} of {{Protocell
  Membranes}}\fg}, \emph{Journal of the American Chemical Society}, 2016,
  \textbf{138}, \cdrnumero 51, 16669-16676.

\bibitem{viricel2015a}
W.~Viricel, A.~Mbarek, J.~Leblond, {\og Switchable {{Lipids}}: {{Conformational
  Change}} for {{Fast pH-Triggered Cytoplasmic Delivery}}\fg}, \emph{Angewandte
  Chemie}, 2015,  \textbf{54}, \cdrnumero 43, 12743-12747.

\bibitem{scouarnec2026}
R.~L. Scouarnec, J.~M. Galindo, C.~Bonduelle, A.~Heise, {\og Polyproline: {{A
  Synthetic Polypeptide With Unique Properties}}\fg}, \emph{Angewandte Chemie
  International Edition},  \textbf{n/a}, \cdrnumero n/a, e5624080.

\bibitem{badreldin2022}
M.~Badreldin, R.~Le~Scouarnec, S.~Lecommandoux, S.~Harrisson, C.~Bonduelle,
  {\og Memory {{Effect}} in {{Thermoresponsive Proline}}-Based
  {{Polymers}}\fg}, \emph{Angewandte Chemie International Edition}, 2022,
  \textbf{61}, \cdrnumero 46, e202209530.

\bibitem{lescouarnec2026}
R.~Le~Scouarnec, E.~Morvan, A.~Gr{\'e}lard, E.~Ibarboure, S.~Lecommandoux,
  C.~Bonduelle, J.~Leblond~Chain, {\og Polyprolinated {{Liposomes}}: {{Toward
  Thermoresponsive Shielding}} and {{Controlled Release}}\fg}, \emph{Langmuir:
  the ACS journal of surfaces and colloids}, 2026,  \textbf{42}, \cdrnumero 28,
  20525-20536.

\bibitem{lescouarnecPHD}
R.~Le~Scouarnec, {\og Lipopolym\`eres Biomim\'etiques et Lipides Stimulables
  Pour Moduler La Perm\'eabilit\'e Membranaire\fg}, These de doctorat,
  Bordeaux, 2025.

\bibitem{equy2025}
E.~Equy, E.~Ibarboure, E.~Grelet, S.~Lecommandoux, {\og Janus {{Polymeric Giant
  Vesicles}} on {{Demand}}: {{A Predictive Phase Separation Approach}} for
  {{Efficient Formation}}\fg}, \emph{Journal of the American Chemical Society},
  2025,  \textbf{147}, \cdrnumero 11, 9727-9738.

\bibitem{manseau2026}
C.~Manseau, {\'E}.~Equy, P.~Patnaik, A.~Mutschler, E.~Ibarboure, T.~P. Lodge,
  S.~Lecommandoux, {\og Programmable {{Budding}} and {{Fission}} of {{Janus
  Polymersomes}} for {{Tunable Cargo Release}}\fg}, \emph{Angewandte Chemie
  International Edition}, 2026, e2878800.

\bibitem{willems2026}
V.~Willems, P.~Moreno, J.~Fojo, L.~{Rodr{\'i}guez-Arco}, L.~Alvarez, {\og
  Life-like Processes in Synthetic Protocells under External Fields\fg},
  \emph{Newton}, 2026,  \textbf{2}, \cdrnumero 7, 100444.

\bibitem{kas1991}
J.~K{\"a}s, E.~Sackmann, {\og Shape Transitions and Shape Stability of Giant
  Phospholipid Vesicles in Pure Water Induced by Area-to-Volume Changes\fg},
  \emph{Biophysical Journal}, 1991,  \textbf{60}, \cdrnumero 4, 825-844.

\bibitem{villalobos-concha2026}
C.~{Villalobos-Concha}, J.~Davis, A.~Baron, L.~Alvarez, {\og Optothermal
  Assembly and Manipulation of Colloids via Non-Coherent Light\fg},
  \emph{Journal of Physics: Photonics}, 2026,  \textbf{8}, \cdrnumero 1,
  015070.

\bibitem{needleman2017}
D.~Needleman, Z.~Dogic, {\og Active Matter at the Interface between Materials
  Science and Cell Biology\fg}, \emph{Nature Reviews Materials}, 2017,
  \textbf{2}, \cdrnumero 9, 17048.

\bibitem{takatori2020}
S.~C. Takatori, A.~Sahu, {\og Active {{Contact Forces Drive Nonequilibrium
  Fluctuations}} in {{Membrane Vesicles}}\fg}, \emph{Physical Review Letters},
  2020,  \textbf{124}, \cdrnumero 15, 158102.

\bibitem{vutukuri2020a}
H.~R. Vutukuri, M.~Hoore, C.~{Abaurrea-Velasco}, L.~Van~Buren, A.~Dutto,
  T.~Auth, D.~A. Fedosov, G.~Gompper, J.~Vermant, {\og Active Particles Induce
  Large Shape Deformations in Giant Lipid Vesicles\fg}, \emph{Nature}, 2020,
  \textbf{586}, \cdrnumero 7827, 52-56.

\bibitem{hayward2006}
R.~C. Hayward, A.~S. Utada, N.~Dan, D.~A. Weitz, {\og Dewetting {{Instability}}
  during the {{Formation}} of {{Polymersomes}} from
  {{Block-Copolymer-Stabilized Double Emulsions}}\fg}, \emph{Langmuir}, 2006,
  \textbf{22}, \cdrnumero 10, 4457-4461.

\bibitem{peyret2018}
A.~Peyret, E.~Ibarboure, J.-F. Le~Meins, S.~Lecommandoux, {\og Asymmetric
  {{Hybrid Polymer}}--{{Lipid Giant Vesicles}} as {{Cell Membrane Mimics}}\fg},
  \emph{Advanced Science}, 2018,  \textbf{5}, \cdrnumero 1, 1700453.

\bibitem{rikken2016}
R.~S.~M. Rikken, H.~Engelkamp, R.~J.~M. Nolte, J.~C. Maan, J.~C.~M. Van~Hest,
  D.~A. Wilson, P.~C.~M. Christianen, {\og Shaping Polymersomes into
  Predictable Morphologies via Out-of-Equilibrium Self-Assembly\fg},
  \emph{Nature Communications}, 2016,  \textbf{7}, \cdrnumero 1, 12606.

\bibitem{gracia2010}
R.~S. Graci{\`a}, N.~Bezlyepkina, R.~L. Knorr, R.~Lipowsky, R.~Dimova, {\og
  Effect of Cholesterol on the Rigidity of Saturated and Unsaturated Membranes:
  Fluctuation and Electrodeformation Analysis of Giant Vesicles\fg}, \emph{Soft
  Matter}, 2010,  \textbf{6}, \cdrnumero 7, 1472.

\bibitem{steinkuhler2019}
J.~Steink{\"u}hler, E.~Sezgin, I.~Urban{\v c}i{\v c}, C.~Eggeling, R.~Dimova,
  {\og Mechanical Properties of Plasma Membrane Vesicles Correlate with Lipid
  Order, Viscosity and Cell Density\fg}, \emph{Communications Biology}, 2019,
  \textbf{2}, \cdrnumero 1, 337.

\bibitem{ashkin1997}
A.~Ashkin, {\og Chapter 1 {{Forces}} of a {{Single-Beam Gradient Laser Trap}}
  on a {{Dielectric Sphere}} in the {{Ray Optics Regime}}\fg}, in \emph{Methods
  in {{Cell Biology}}}, Methods in {{Cell Biology}}, vol.~55, Elsevier, 1997,
  1-27.

\bibitem{willems2025}
V.~Willems, A.~Baron, D.~{Fernandez-Matoz}, G.~Wolfisberg, J.-C. Baret,
  E.~Dufresne, L.~Alvarez, {\og Run-and-Tumble Dynamics of Active Giant
  Vesicles\fg}, \emph{Soft Matter}, 2025,  \textbf{21}, \cdrnumero 31,
  6175-6185.

\bibitem{Luisi2006}
P.~L. Luisi, F.~Ferri, P.~Stano, {\og Approaches to Semi-Synthetic Minimal
  Cells: A Review\fg}, \emph{Die Naturwissenschaften}, 2006,  \textbf{93},
  \cdrnumero 1, 1-13.

\bibitem{lartigue2009a}
C.~Lartigue, S.~Vashee, M.~A. Algire, R.-Y. Chuang, G.~A. Benders, L.~Ma, V.~N.
  Noskov, E.~A. Denisova, D.~G. Gibson, N.~{Assad-Garcia}, N.~Alperovich, D.~W.
  Thomas, C.~Merryman, C.~A. Hutchison, H.~O. Smith, J.~C. Venter, J.~I. Glass,
  {\og Creating Bacterial Strains from Genomes That Have Been Cloned and
  Engineered in Yeast\fg}, \emph{Science}, 2009,  \textbf{325}, \cdrnumero
  5948, 1693-1696.

\bibitem{labroussaa2016a}
F.~Labroussaa, A.~Lebaudy, V.~Baby, G.~Gourgues, D.~Matteau, S.~Vashee,
  P.~{Sirand-Pugnet}, S.~Rodrigue, C.~Lartigue, {\og Impact of Donor-Recipient
  Phylogenetic Distance on Bacterial Genome Transplantation\fg}, \emph{Nucleic
  Acids Research}, 2016,  \textbf{44}, \cdrnumero 17, 8501-8511.

\bibitem{JittaseviPHD}
{\og D\'ecryptage Des M\'ecanismes de La Transplantation de G\'enomes Entiers
  Chez Les {{Mollicutes}} \textbar{} {{Theses}}.Fr\fg},
  https://theses.fr/2025BORD0466.

\bibitem{moragues2023}
T.~Moragues, D.~Arguijo, T.~Beneyton, C.~Modavi, K.~Simutis, A.~R. Abate, J.-C.
  Baret, A.~J. {deMello}, D.~Densmore, A.~D. Griffiths, {\og Droplet-Based
  Microfluidics\fg}, \emph{Nature Reviews Methods Primers}, 2023,  \textbf{3},
  \cdrnumero 1, 32.

\bibitem{beneyton2020b}
T.~Beneyton, C.~Love, M.~Girault, T.-Y.~D. Tang, J.-C. Baret, {\og
  High-{{Throughput Synthesis}} and {{Screening}} of {{Functional Coacervates
  Using Microfluidics}}\fg}, \emph{ChemSystemsChem}, 2020,  \textbf{2},
  \cdrnumero 6, e2000022.

\bibitem{weiss2018d}
M.~Weiss, J.~P. Frohnmayer, L.~T. Benk, B.~Haller, J.-W. Janiesch, T.~Heitkamp,
  M.~B{\"o}rsch, R.~B. Lira, R.~Dimova, R.~Lipowsky, E.~Bodenschatz, J.-C.
  Baret, T.~{Vidakovic-Koch}, K.~Sundmacher, I.~Platzman, J.~P. Spatz, {\og
  Sequential Bottom-up Assembly of Mechanically Stabilized Synthetic Cells by
  Microfluidics\fg}, \emph{Nature Materials}, 2018,  \textbf{17}, \cdrnumero 1,
  89-96.

\bibitem{shang2017}
L.~Shang, Y.~Cheng, Y.~Zhao, {\og Emerging {{Droplet Microfluidics}}\fg},
  \emph{Chemical Reviews}, 2017,  \textbf{117}, \cdrnumero 12, 7964-8040.

\bibitem{breukers2025}
J.~Breukers, K.~Ven, W.~Verbist, I.~Rutten, J.~Lammertyn, {\og From
  Specialization to Broad Adoption: {{Key}} Trends in Droplet Microfluidic
  Innovations Enhancing Accessibility to Non-Experts\fg},
  \emph{Biomicrofluidics}, 2025,  \textbf{19}, \cdrnumero 2, 021302.

\bibitem{abrishamkar2026}
A.~Abrishamkar, L.~Neely, C.~Pham, N.~Ibrahim, A.~Larios-Cervantes, B.~A.
  Kertesz, T.~Hoare, {\og Microfluidic {{Engineering}} of {{Microgels}}: {{A
  Review}} of {{Emerging Technologies}}, {{Materials}}, {{Fabrication
  Strategies}}, and {{Applications}}\fg}, \emph{Advanced Healthcare Materials},
  2026,  \textbf{15}, \cdrnumero 41, e71809.

\bibitem{lenzi2026}
E.~Lenzi, N.~Vertti-Quintero, J.~Husson, A.-L. Bulteau, C.~Nizard, K.~Pays,
  S.~Sart, C.~N. Baroud, {\og Spherical {{Skin Model}}: {{Stratified
  Co}}-{{Culture}} of {{Fibroblasts}} and {{Keratinocytes}} on {{Spherical
  Beads Toward Compound Screening}}\fg}, \emph{Advanced Healthcare Materials},
  2026,  \textbf{15}, \cdrnumero 13, e03250.

\bibitem{alessandri2013a}
K.~Alessandri, B.~R. Sarangi, V.~V. Gurchenkov, B.~Sinha, T.~R. Kie{\ss}ling,
  L.~Fetler, F.~Rico, S.~Scheuring, C.~Lamaze, A.~Simon, S.~Geraldo,
  D.~Vignjevi{\'c}, H.~Dom{\'e}jean, L.~Rolland, A.~Funfak, J.~Bibette,
  N.~Bremond, P.~Nassoy, {\og Cellular Capsules as a Tool for Multicellular
  Spheroid Production and for Investigating the Mechanics of Tumor Progression
  in Vitro\fg}, \emph{Proceedings of the National Academy of Sciences}, 2013,
  \textbf{110}, \cdrnumero 37, 14843-14848.

\bibitem{zhao2023}
Y.~Zhao, E.~Y. Wang, F.~B. Lai, K.~Cheung, M.~Radisic, {\og Organs-on-a-Chip: A
  Union of Tissue Engineering and Microfabrication\fg}, \emph{Trends in
  Biotechnology}, 2023,  \textbf{41}, \cdrnumero 3, 410-424.

\bibitem{dai2021}
M.~Dai, J.-P. Bela{\"i}di, G.~Fleury, E.~Garanger, M.~Rielland, X.~Schultze,
  S.~Lecommandoux, {\og Elastin-like {{Polypeptide-Based Bioink}}: {{A
  Promising Alternative}} for {{3D Bioprinting}}\fg}, \emph{Biomacromolecules},
  2021,  \textbf{22}, \cdrnumero 12, 4956-4966.

\bibitem{hunt2025}
A.~C. Hunt, B.~J. Rasor, K.~Seki, H.~M. Ekas, K.~F. Warfel, A.~S. Karim, M.~C.
  Jewett, {\og Cell-{{Free Gene Expression}}: {{Methods}} and
  {{Applications}}\fg}, \emph{Chemical Reviews}, 2025,  \textbf{125},
  \cdrnumero 1, 91-149.

\bibitem{Sierra2026}
A.~M. Restrepo~Sierra, F.~Ramirez~Gomez, M.~{van Tongeren}, L.~Sierra~Heras,
  C.~Danelon, {\og A Synthetic Cell with Integrated {{DNA}} Self-Replication
  and Lipid Biosynthesis\fg}, \emph{Nature Communications}, 2026,  \textbf{17},
  \cdrnumero 1, 2727.

\bibitem{meier2026}
M.~Meier, S.~A. Scholz, M.~Schaaf, N.~Safronova, N.~F. Greenwood, S.~Si,
  I.~Lieberwirth, A.~L. Jung, K.~Landfester, A.~J.~M. Driessen, T.~J. Erb, {\og
  Functional Mapping and Engineering of the {{Sec}} Translocon Unlocked by a
  Cell-Free System\fg}, \emph{Science Advances}, 2026.

\end{thebibliography}

\begin{table*}[!ht]
    \rotatebox{90}{
\begin{tabular}{p{6cm}p{7cm}p{3.5cm}}
Principal Investigator & Funding & Grant number\\
\hline
\hline
J.-C. Baret, N. Martin, L. Alvarez   & EU MSCA DN SigSynCell              & 101119961\\
N. Martin                            & EU MSCA Fellowship, SynCellCluster & 101153428\\
V. Desvergnes                        & EU MSCA Fellowship, BioCataDrops   & 101152804\\
A. Aubret                            & EU MSCA Fellowship, NanoscAM   & 886024\\
T. Salez                             & EU ERC CoG EmetBrown               & 101039103\\
\hline
\hline
M. Aznauryan                   & ANR JCJC DISTINCT      & ANR-20-CE11-0007\\
N. Martin    	               & ANR JCJC LASCO2	& ANR-21-CE06-0022\\
L. Alvarez                     & ANR JCJC MIMESYS       & ANR-23-CE06-0007\\
A. Aubret                      & ANR JCJC NANODART	& ANR-24-CE30-6855\\
\hline
S. Lecommandoux 	       & ANR PRC TEPEE          & ANR-19-CE18-0024\\
J. Leng, J. Amédée             & ANR PRC NEUROSKIN	& ANR-21-CE14-0027\\
L. Béven, N. Martin            & ANR PRC WALLESSHAPE	& ANR-22-CE44-0015\\
S. Lecommandoux, E. Grelet     & ANR PRC POPART         & ANR-23-CE06-0008\\
C. Bonduelle, N. Martin        & ANR PRC PROTOPOLYLM	& ANR-24-CE07-1479\\
D. Coudreuse	               & ANR PRC INTRAFLUIDICS  & ANR-24-CE13-2733\\
S. Saupe, A. Loquet            & ANR PRC HETNA          & ANR-24-CE13-6231\\
D. Coudreuse	               & ANR PRC CROWDPHASE     & ANR-24-CE30-5426\\
\hline
L. Alvarez                     & ANR PRCI APPOLLO	& ANR-23-CE06-0035\\
\hline
C. Lartigue                    & ANR ME EXEGESYS        & ANR-25-CE44-4299\\
\hline
\hline
D. Coudreuse                   & Région Nouvelle Aquitaine CHESS Program & 15963520; 15964420\\
N. Martin, E. Ducrot           & Région Nouvelle Aquitaine               & 2020-2019-8330510\\
J.-C. Baret                    & Fondation Simone et Cino Del Duca 2021  & NA \\
\hline
\hline
\end{tabular}
}
\caption{\textbf{List of third-party funding} which contributed to the progress of the research reported in the current manuscript.} \label{table:2}
\end{table*}

\end{document}